\documentclass[12pt,twocolumn]{emulateapj}

\tightenlines

\usepackage{ulem} 
\usepackage{color} 
\usepackage{graphicx}
\usepackage{mediabb}
\usepackage{amsmath}
\usepackage{times}
\usepackage{rotating}

\makeatletter

\newcommand{\Rmnum}[1]{\expandafter\@slowromancap\romannumeral #1@}
\makeatother

\newcount\listnorom
\newcommand\listromanDE{\global\advance \listnorom by 1
{\lowercase\expandafter{(\romannumeral\listnorom)}\ }}

\def\lsim{\raise0.3ex
  \hbox{$<$\kern-0.75em\raise-1.1ex\hbox{$\sim$}}\,}
\def\gsim{\raise0.3ex
  \hbox{$>$\kern-0.75em\raise-1.1ex\hbox{$\sim$}}\,}

\newcommand{\be}{\begin{eqnarray}}
\newcommand{\ee}{\end{eqnarray}}

\newcount\Inum
\newcount\IInum
\newcount\IIInum
\newcount\IVnum
\def\I{\global\multiply\IInum by 0 \global\multiply\IIInum by 0
            \global\multiply\IVnum by 0 \global\advance \Inum by 1
            {\the\Inum. }}
\def\II{\global\multiply\IIInum by 0\global\multiply\IVnum by 0
       \global\advance \IInum by 1 {\the\Inum.\the\IInum. }}
\def\III{\global\multiply\IVnum by 0\global\advance \IIInum by 1
            {\the\Inum.\the\IInum.\the\IIInum. }}
\def\IV{\global\advance \IVnum by 1
            {\the\IVnum. }}
\begin{document}

\title{Time evolution of broadband non-thermal emission from supernova remnants in different circumstellar environments}

\author{
Haruo Yasuda \altaffilmark{1}
Shiu-Hang Lee \altaffilmark{1}
}

\altaffiltext{1}{Department of Astronomy, Kyoto University, Kitashirakawa, Oiwake-cho, Sakyo-ku, Kyoto 606-8502, Japan; yasuda@kusastro.kyoto-u.ac.jp}

\begin{abstract}
Supernova remnants (SNRs) are thought to be one of the major acceleration sites of galactic cosmic rays (CRs) and an important class of objects for high-energy astrophysics. SNRs produce multi-wavelength, non-thermal emission via accelerated particles at collisionless shocks generated by the interactions between the SN ejecta and the circumstellar medium (CSM). Although it is expected that the rich diversities observed in supernovae (SNe) and their CSM can result in distinct very-high-energy (VHE) electromagnetic signals in the SNR phase, there are only a handful of SNRs observed in both GeV and TeV $\gamma$-rays so far. A systematic understanding of particle acceleration at SNRs in different ambient environments is therefore limited. Here, we explore non-thermal emission from SNRs in various circumstellar environments up to 5000 yrs from explosion using hydrodynamical simulations coupled with efficient particle acceleration. We find that time-evolution of emission characteristics in the VHE regime is mainly dictated by two factors; the number density of the target particles and the amplified magnetic field in the shocked medium. We also predict that Cherenkov Telescope Array (CTA) will have a sufficient sensitivity to detect VHE $\gamma$-rays from most young SNRs at distances $\lsim 5.0\ \mathrm{kpc}$. Future SNR observations with CTA will thus be promising for probing the CSM environment of SNe and hence their progenitor properties, including the mass loss history of massive stars.
\end{abstract}


\section{Introduction}
 Since \cite{1934PNAS...20..259B} suggested the relation between supernovae (SNe) and cosmic-rays (CRs), supernova remnants (SNRs) have been studied as the accelerators of Galactic CRs below the ``knee'' energy ($\sim 3\mathrm{PeV}$). One of the most successful theories currently for the particle acceleration mechanism is the so-called diffusive shock acceleration (DSA) \citep[e.g.,][]{1949PhRv...75.1169F,1983SSRv...36...57D,2010APh....33..307C,2010MNRAS.407.1773C} which has been widely studied in the last couple of decades, nevertheless there still remains much to be understood in the details of the microphysical processes.

SNRs are commonly detected in multi-wavelength observstions and some have been found to shine in a broad range of frequencies from radio all the way to TeV $\gamma$-rays. In general they emit broadband non-thermal electromagnetic radiation due to their interactions with the interstellar matter (ISM) or circumstellar medium (CSM). The radio and non-thermal X-rays are believed to be produced by relativistic electrons through synchrotron radiation. The $\gamma$-rays can originate from both relativistic electrons through inverse Compton scattering (IC) and bremsstrahlung, as well as by relativistic protons through the $\pi^0$-decay channel from proton-proton inelastic scatterings, which are usually regarded as the {\it leptonic} and {\it hadronic} processes, respectively. 

\begin{figure}
 \centering
 \includegraphics[width=8cm]{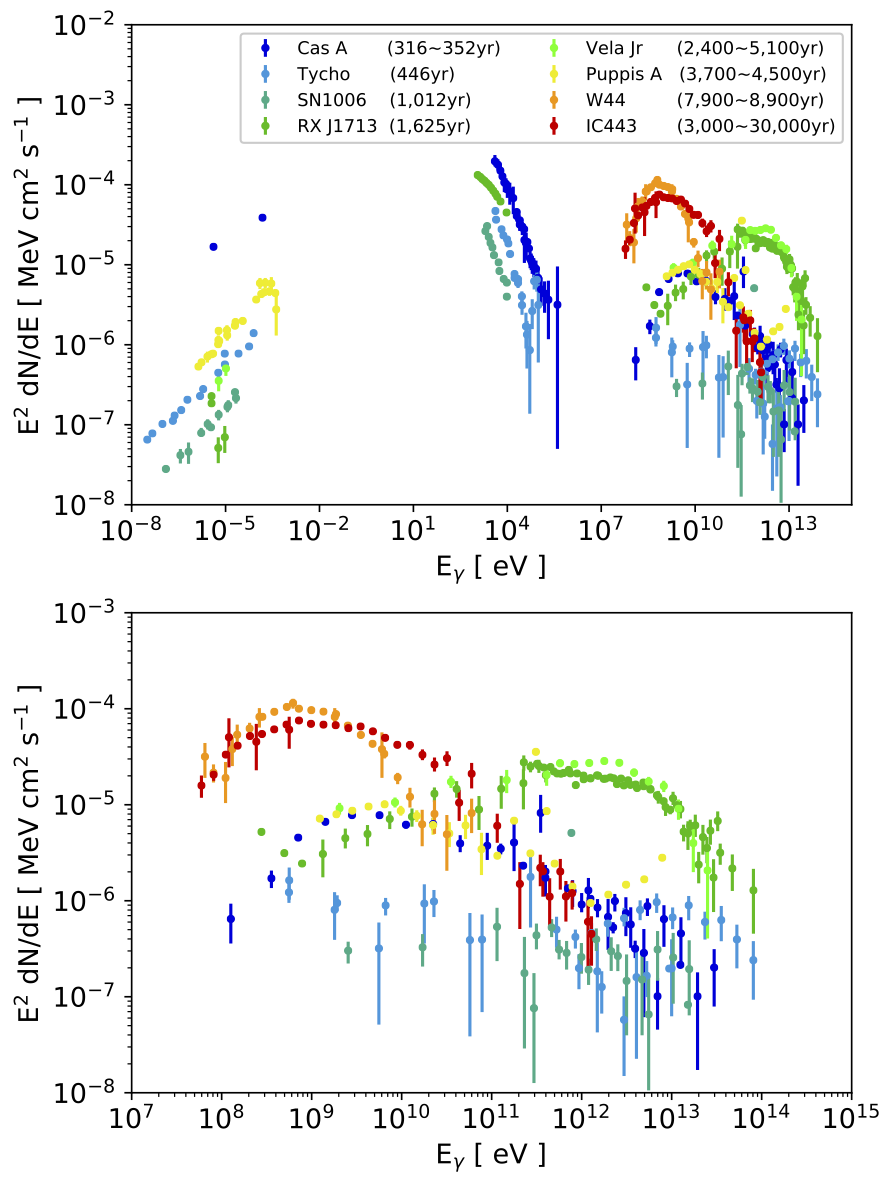}
 \caption{Upper panel: multi-wavelength SED of the SNRs whose $\gamma$-ray flux is detected. The color of data points almost represents the SNR age, the redder the color become, the older the age of SNRs become. Lower panel: the same plot as the upper panel, but energy range is from 10 MeV to 1 PeV.}
 \label{graph:SED} 
\end{figure}

Fig.~\ref{graph:SED} shows the spectral energy distribution (SED) of SNRs which have been observed so far in the GeV to TeV energy range. The upper panel shows the overall SED from radio to 1~PeV, and the lower panel shows the $\gamma$-ray SED from 10~MeV to 1~PeV. In most cases, the radio and non-thermal X-ray spectrum can be satisfactorily reproduced by a synchrotron origin regardless of SNR age, but the differences in the observed $\gamma$-ray spectra among these SNRs are remarkable. 
Whether the $\gamma$-rays are produced by either {\it hadronic} or {\it leptonic} (or both) channel has a large implication on the particle acceleration mechanism, such as the injection efficiencies of the supra-thermal particles, the maximum energy of the accelerated particles, and the overall acceleration efficiency. These aspects can vary significantly among different individual SNRs depending on their ambient environment, age and progenitor system which need to be fully understood in a consistent picture in order to examine the SNR population as a dominant source of Galactic CRs. However, the model interpretation is still often found to be controversial and remains to be a subject for discussion.   

A general picture has been proposed by recent works \citep[e.g.,][]{2012ApJ...761..133Y} that the observed properties of the $\gamma$-ray emission from SNRs are mainly determined by the gas density in the their surrounding environments, i.e., the dominant component of the $\gamma$-ray flux is IC if the SN occurred in a relatively tenuous medium, while the $\pi^0$-decay component dominates in a denser medium such as a molecular cloud.These results, however, are usually based on phenomenological fitting of the observed photon SED from individual SNRs using simple one-zone models. From the theoretical point of view, previous works \citep[e.g.,][]{2008MNRAS.384.1119F,2016ApJS..227...28T,2018MNRAS.475.5237G} also follow the long-term time-evolution of broadband emission, but assumptions and simplifications like one-zone hydrodynamical model and simple power-law CR spectrum are usually employed in these calculations. Currently, there are few studies that follow the long-term time evolution of the broadband emission together with the hydrodynamical evolution of the SNRs coupled to a self-consistent treatment of DSA at the shocks. Here, using a multi-zone hydrodynamical simulation coupled with an efficient particle acceleration, we generate a grid of evolutionary models of SNRs interacting with various kinds of ISM/CSM environments up to a few 1000 yrs over an observation-based parameter space. Our results are analyzed to explore general trends in the characteristics of the time-evolving SED that can be used in the future as a probe of the structure of the surrounding environment. Based on our results, we also predict the observability of typical young core-collapse and Type Ia SNRs by the upcoming ground-based VHE $\gamma$-ray observatory Cherenkov Telescope Array (CTA) . 

In section~\ref{Method}, we introduce our numerical method for the evolutionary model of SNRs and the range of models adopted for the ambient environment. Section~\ref{Results} describes our results and interpretations from the calculation, including the time-evolution of the SNR dynamics and the multi-wavelength spectra, and comparison to the currently available observational data. Concluding remarks and summary can be found in Section~\ref{Conclusion}.

\section{Method}\label{Method}
\subsection{Simulation code and included physics}\label{physics}
We develop a hydrodynamical code to investigate the effect of CSM interaction on the long-term evolution of non-thermal radiations from SNRs. The code performs 1-D spherically symmetric hydro simulations on a Lagrangian mesh based on the VH-1 code \citep[e.g.,][]{2001ApJ...560..244B} coupled with a semi-analytic non-linear DSA (NLDSA) calculation \citep[see, e.g.,][]{2004APh....21...45B,2010APh....33..307C,2010MNRAS.407.1773C} similar to the framework first introduced in the {\it CR-hydro-NEI} code \citep[see, e.g.,][]{LEN2012}. The time-evolution of the SNR is numerically calculated through a Lagrangian hydrodynamics simulation starting from a self-similar model for the SN ejecta as initial condition. The expansion of the SNR into whatever ambient environment adopted in a model is then followed by the hydro simulation, from which the shock dynamics is traced in real-time as an input for a NLDSA calculation. The NLDSA part provides a solution for the accelerated CR which feedbacks to the hydrodynamics through an effective gamma approach, i.e., a modified equation-of- state in the shocked medium \citep{2001ApJ...560..244B}, as well as the occurrence of a shock precursor. 

NLDSA is sensitive to the shock velocity and the gas density and the magnetic field strength in the upstream environment, so we improve their code to calculate the DSA process and its hydrodynamical feedbacks at the shock every time the shock sweeps up gas in a new (unshocked) grid. This is particularly important in the case of a structured ambient medium such as a confined CSM due to an episodic mass loss from a massive star (see, Section~\ref{sub:en}). 

In the NLDSA calculation, we obtain the phase-space distribution function $f(x,p)$ of the accelerated protons by solving the following diffusion-convection equation written in the shock-rest frame \citep[e.g.,][]{2010APh....33..307C,2010MNRAS.407.1773C,LEN2012}, assuming a steady-state \footnote{We consider that it is reasonable to use the steady-state approximation as long as the dynamical time-scale of the SNR is longer than the DSA acceleration time-scale $t_\mathrm{acc}$. Known young SNRs are found to accelerate protons up to a maximum momentum $\sim$100 TeV/c or below, so that $t_\mathrm{acc}\sim D/u^2\sim1(p/100\ \mathrm{TeV/c})(B/100\ \mu\mathrm{G})^{-1}(u/3000\ \mathrm{km/s})^{-2}\ \mathrm{yr}$, where $D,\ u,\ p,\ B$ are the diffusion coefficient, the shock velocity, the particle momentum, and the amplified magnetic field. We find that the above condition can be satisfied within the scope of our models.\label{fn:steady}} distribution isotropic in momentum space,
\begin{eqnarray}\label{eq:DSA}
&&[u(x)-v_A(x)]\frac{\partial f(x,p)}{\partial x}-Q(x,p)= \notag\\
&&\frac{\partial}{\partial x}\left[D(x,p)\frac{\partial f(x,p)}{\partial x}\right]+\frac{p}{3}\frac{d[u(x)-v_A(x)]}{dx}\frac{\partial f(x,p)}{\partial p} ,
\end{eqnarray}
where $D(x,p), u(x), v_A(x)$ are the spatial diffusion coefficient, gas velocity and Alfv\'en speed in the shock-rest frame at each position $x$. Hereafter, we label each quantity with a subscript `0', `1', and `2' denoting values at far upstream ($x=-\infty$), immediately upstream ($x=0^-$), and immediately downstream ($x=0^+$) from the shock, respectively. We assume a Bohm diffusion for the accelerating particles in this work, such that $D(x,p)=pc^2/3eB(x)$, where $B(x)$ is the local magnetic field strength at position $x$. The magnetic field is self-consistently calculated with magnetic field amplification (MFA) due to self-generated turbulence through resonant CR streaming instability \citep[e.g.,][]{1978MNRAS.182..147B,CBAV2009}. Following \cite{2004APh....21...45B} and \cite{2005MNRAS.361..907B}, we adopt the `thermal-leakage' injection model for the DSA injection rate $Q(x,p)$ such that 
\begin{equation}
Q(x,p) = \eta\frac{n_1u_1}{4\pi p_\mathrm{inj}^2}\delta(x)\delta(p-p_\mathrm{inj}),
\end{equation}
where $\eta=\{4/(3\sqrt{\pi})\}(S_\mathrm{sub}-1)\chi_\mathrm{inj}^3e^{-\chi_\mathrm{inj}^2}$ and  $S_\mathrm{sub}=(u_1-v_{A,1})/(u_2+v_{A,2})$ is the effective compression ratio that the streaming particles experience at the sub-shock position ($x=0$). $p_\mathrm{inj} \equiv \chi_\mathrm{inj}\sqrt{2m_\mathrm{p}k_\mathrm{b}T_\mathrm{p}}$ is the DSA injection momentum, where $m_\mathrm{p}=1.6\times 10^{-24}\ \mathrm{g}$ is the mass of the proton, $T_\mathrm{p}$ is the proton temperature, and $\chi_\mathrm{inj}$ is a free parameter constrained by observations. By solving eq.~(\ref{eq:DSA}), The distribution function at the shock position with a cutoff at a maximum momentum $p_\mathrm{max}$ can be written in implicit form as below,
\begin{eqnarray}\label{eq:fp}
f_1(p)=&&f(x=0,p)\notag\\
=&&\frac{3S_\mathrm{tot}}{S_\mathrm{tot}U(p)-1}\times\notag\\
&&\left[\frac{\eta n_0}{4\pi p_\mathrm{inj}^3}\exp\left(-\int_{p_\mathrm{inj}}^p \frac{dp'}{p'}\frac{3S_\mathrm{tot}U(p')}{S_\mathrm{tot}U(p')-1}\right)\right]\times \notag\\
&&\qquad\qquad\qquad\qquad\qquad \exp\left\{-\left(\frac{p}{p_\mathrm{max}}\right)^\alpha\right\},
\end{eqnarray}
where $S_\mathrm{tot}=(u_0-v_{A,0})/(u_2+v_{A,2})$ is the effective total compression ratio of the CR-modified shock. The explicit expressions of $u(x),\ v_A(x),\ U(p)$ can be found in \citet[][]{LEN2012} and reference therein. The parameter $\alpha$ describes the rollover shape near the high-energy cutoff which serves as a parametrization of the poorly understood escape process of the accelerated particles.

As for the electrons whose gyroradii are much smaller at thermal energies, the injection mechanism and efficiency relative to their proton and ion counterparts at strong collisionless shocks are still not fully understood, although a few first-principle kinetic simulations have shed new light onto this topic recently \citep[see, e.g.,][and reference therein]{PhysRevLett.119.105101}. In this work, we constrain the electron-to-proton number ratio ($K_\mathrm{ep}$) at relativistic momenta below the cutoff by currently available data from multi-wavelength observations. Current observations of young $\gamma$-ray emitting SNRs have constrained $K_\mathrm{ep}$ to a range of a few $10^{-4}$ to $\sim 10^{-2}$ \citep[e.g.,][]{2018AA...612A...6H}. In this work, we adopt a $K_\mathrm{ep}$ by performing calibrations against data from prototypical Type Ia and core-collapse (CC) SNRs (see Section~\ref{calibrate}).

The maximum momentum of the accelerated protons is constrained by a number of physical conditions as described below, and its value is taken to be the minimum of the momenta obtained by applying these conditions, i.e., $p_\mathrm{max,p}=\min\{p_\mathrm{max,age},p_\mathrm{max,feb}\}$, which changes with time as the shock propagates and evolves. The condition for $p_\mathrm{max,age}$ ({\it age-limited}) comes from the comparison of the SNR age $t_\mathrm{age}$ with the DSA acceleration time-scale $t_\mathrm{acc}$.
An approximate expression for $t_\mathrm{acc}$ can be written as
\begin{equation}
t_\mathrm{acc}\approx \frac{3}{u_0-u_2}\int^{p_\mathrm{max}}_{p_\mathrm{inj}}\frac{dp}{p}\left(\frac{D_0(p)}{u_0}+\frac{D_2(p)}{u_2}\right) ,
\end{equation}
where $D_0(p)\ (D_2(p))$ is the diffusion coefficient at far upstream (immediate downstream) from the shock.

The condition for $p_\mathrm{max,feb}$ ({\it escape-limited}) comes from the spatial confinement of the accelerating particles, i.e., a comparison of the particle diffusion length $L_\mathrm{diff}$ with a free-escape-boundary (FEB) set at a distance $L_\mathrm{feb}$ upstream from the sub-shock. Here $L_\mathrm{feb} = f_\mathrm{feb}R_\mathrm{sk}$ where $f_\mathrm{feb}$ is typically taken between 0.1 and 0.2 motivated by currently available models of SNR observations \citep[e.g.,][]{2009MNRAS.396.2065C,LEN2012}. We fix $f_\mathrm{feb}$ at 0.1 in this study \footnote{ In our models, $p_\mathrm{max}$ is typically constrained by {\it age-limited} for a $t_\mathrm{age} \le 100\ \mathrm{yr}$, and then becomes {\it escape-limited} afterwards. The exact timing of the transition depends on the ambient medium in which the SN ejecta expands into in the early phase (see Fig.~\ref{out:n} and \ref{out:m}). }. $L_\mathrm{diff}$ can be obtained by the following expression, 
\begin{equation}
L_\mathrm{diff} = \left< \frac{D(x, p_\mathrm{max})}{u(x)}\right> = \int^0_{-L_\mathrm{feb}}\frac{dx}{L_\mathrm{feb}}\frac{D_0(x, p_\mathrm{max})}{u(x)}.
\end{equation}

For electrons, $p_\mathrm{max,e}$ is further restricted by the efficient energy loss due to radiation ({\it loss-limited}), that is $p_\mathrm{max,e}=\min\{p_\mathrm{max,age},p_\mathrm{max,feb},p_\mathrm{max,loss}\}$. The condition for $p_\mathrm{max,loss}$ ({\it loss-limited}) derives from the comparison of the acceleration time-scale $t_\mathrm{acc}$ with the time-scale of energy losses from non-thermal emission $t_\mathrm{loss}$. 
Typically, synchrotron radiation and IC dominate the energy loss of relativistic electrons, hence we can obtain $t_\mathrm{loss}$ as follows,
\begin{equation}
t_\mathrm{loss} = \frac{3m_\mathrm{e}c^2}{4c\sigma_\mathrm{T}U_\mathrm{B,2}\gamma}\left(1+\sum_{i=1}^{N_\mathrm{p}}\frac{W_i\gamma_{k,i}^2}{U_\mathrm{B,2}(\gamma^2+\gamma_{k,i}^2)}\right)^{-1}, 
\end{equation}  
where $U_\mathrm{B,2} = {B_2}^2/8\pi$ is magnetic field energy density in downstream and $\sigma_\mathrm{T}, \gamma$ are the Thompson cross section and electron Lorentz factor respectively. $N_\mathrm{p}$ is the number of components of external photon fields and $\gamma_{k,i} = 0.53m_\mathrm{e}c^2/k_\mathrm{b}T_i$ is the critical Lorentz factor. $W_i, T_i$ are the energy density and effective temperature of the $i$-th component of the seed photon fields. In this study, we only consider the cosmic microwave background radiation (CMB) as the target photons for simplicity, so $W_i$ = 0.26 eV\ cm$^{-3}$ and $T_i$ = 2.7 K.

After the particles are accelerated at the shock, they advect with the gas flow in the downstream region assuming an effective trapping by the strong, amplified magnetic turbulence. During the advection, they lose energy in the meantime through non-thermal emission due to interactions with the shocked ISM/CSM, and the adiabatic expansion of the SNR. Following \cite{1997ApJ...490..619S}, the radiation loss mechanisms include synchrotron radiation, bremsstrahlung, IC for electrons, and an addition of pion productions for protons. Coulomb loss is not included in this work but can be important for sub-GeV $\gamma$-ray emission.

Using the calculated proton and electron spectra in each position at any given age, we can then calculate the broadband non-thermal emission spectra. Our code includes synchrotron, IC, thermal and non-thermal bremsstrahlung and $\pi^0$-decay emission by the accelerated particles, taking also into account the additional contributions from secondary electrons and positrons on the synchrotron, IC and non-thermal bremsstrahlung components. We apply eq. (D1)-(D7) in \cite{2010PhRvD..82d3002A} to calculate the volume emissivities for synchrotron radiation, eq. (29)-(33) in \cite{1997ApJ...490..619S} for IC, eq. (26)-(28) in \cite{1997ApJ...490..619S} for non-thermal electron-proton bremsstrahlung, eq. (A1)-(A7) in \cite{1999ApJ...513..311B} for non-thermal electron-electron bremsstrahlung, and the parametrized model presented in \cite{2006ApJ...647..692K} for the $\pi^0$-decay $\gamma$-ray emission. The code also computes the thermal bremsstrahlung emission using eq. (5.14) in \cite{1986rpa..book.....R}. For this component, we assumed that the shocked gases are fully ionized after shock-heating so that the electron number density is $n_\mathrm{e}(x) = (1 + f_\mathrm{He}) \times \rho(x)/\mu m_\mathrm{p}$, where $\mu=1.4$ is the mean molecular weight assuming the number fraction of helium $f_\mathrm{He}$ is 10$\%$ of hydrogen in the ambient medium. 

The shock-heated proton and electron temperatures are assumed to be proportional to the mass number for a collisionless shock, and they are further evolved in the downstream due to adiabatic cooling/heating and equilibration through coulomb collisions \citep[i.e. eq. 5-31 in ][]{1965pfig.book.....S}. We also include free-free absorption and synchrotron-self absorption with eq. (5.18) and (6.50) in \cite{1986rpa..book.....R} which are important in the radio band. The treatment of secondary electron/positron production through $\pi^{+/-}$ decay and subsequent photon emission follows the method described in  \cite{2015ApJ...806...71L}.

After the SNR has entered the radiative phase, the shock slows down to an extent that DSA is expected to be inefficient relative to the younger stages (see, however, \citet{2015ApJ...806...71L} and reference therein for a discussion on GeV-bright middle-aged SNRs). We do not treat the physics involved in radiative shocks in this work, and the simulations are terminated before the SNR becomes radiative. For all cases, we run the models up to an age of 5000 yr which is still within the Sedov-Taylor phase. We also do not consider the acceleration of heavy ions and possible DSA at the reverse shock in this study. These aspects will be discussed in future works.

\subsection{Models for the surrounding environment }\label{code:model}
In this study, we look at two classes of simple but representative models for the ambient medium around a SNR. In Model A and its variants, we consider a uniform ISM-like environment which is usually expected for a Type Ia SNR (with exceptions),
\begin{eqnarray}
 \rho(r) &=& \mu m_\mathrm{p} n_\mathrm{ISM}, \\
 B(r)     &=& B_0,
\end{eqnarray}
where $n_\mathrm{ISM},\ B_0$ are the ISM proton number density and magnetic field. We use an exponential profile for the SN ejecta in these models \citep{1998ApJ...497..807D}.

Model B and its variants adopt a power-law spatial distribution for the density in the ambient gas, which mimics the CSM created by a non-episodic isotropic stellar wind from a massive star prior to core-collapse supernova (CCSN) \citep[][and references therein]{2012ApJ...744...39E},   
\begin{eqnarray}
 \rho(r) &=& \frac{\dot{M}}{4\pi V_\mathrm{w}}r^{-2}, \\
 B(r)     &=& \frac{(\sigma_\mathrm{w} V_\mathrm{w} \dot{M})^{1/2}}{r},
\end{eqnarray}
where $\dot{M},\ V_\mathrm{w},\ \sigma_\mathrm{w}$ are the mass loss rate, wind velocity and the ratio between the magnetic field energy density and the wind kinetic energy density, i.e., $\sigma_\mathrm{w} \equiv P_\mathrm{B}/E_{\mathrm{kin,w}}=(B^2/8\pi)/(\rho V_\mathrm{w}^2/2)$.
We use a flat core with power-law envelope profile for the ejecta in these CCSN-like models \citep{1999ApJS..120..299T}. In both classes of models, we assume that the gas velocity and temperature of the unshocked material are constant in space.

In Model C, we investigate the case of a non-steady mass loss history from a massive star in which a dense shell (or confined CSM) surrounding the ejecta is created due to mass ejection from the stellar envelope during the course of a few hundred years before the CC onset. The CSM is represented by a simple combination of two wind profiles where the one in the inner region having a higher density, as below,
\begin{eqnarray}
 \rho(r) &=& \frac{\dot{M}}{4\pi V_\mathrm{w}}r^{-2}\ \ \ \ \ \ \ \ \ \ \ \ \ \ \ \ \ \ \ \ \ \ \ \ \ \ \ (R_\mathrm{tr}<r) \notag \\
            &=& \frac{\dot{M_2}}{4\pi V_{\mathrm{w},2} R_\mathrm{tr}^2}\left(\frac{r}{R_\mathrm{tr}}\right)^{-n_\mathrm{pl,2}}\ \ (r\le R_\mathrm{tr}),
\end{eqnarray}
where $\dot{M_2},V_\mathrm{w,2},n_\mathrm{pl,2}$ are the mass loss rate, velocity and power-law index of the wind profile from an enhanced mass loss, and $R_\mathrm{tr}$ is the transition radius between the normal wind and the confined CSM region. As typical values, we consider an episode that an enhanced mass loss ejection with $V_{w,2}\sim\ $1000 km/s occurred during the last $\sim\ $1000 yr before explosion, and $R_\mathrm{tr} \sim\ $1.0 pc. 

\section{Results and Discussion}\label{Results}

\subsection{Calibration models}\label{calibrate}
To cross-check the robustness of the code and its capability of reproducing observations, 
we first consider two models, A0 and B0, with parameters chosen to match the multi-wavelength observation data of the Ia SNR Tycho and the CC SNR RX~J1713.7-3946 (hereafter RX~J1713) based on the multi-wavelength emission model from previous hydro simulations presented in \citet{Slane2014} and \citet{LEN2012}, respectively. Model A and B and their variants will then be generated based on these observationally calibrated models by varying the ambient environment.

\begin{figure}
\centering
\includegraphics[width=8cm]{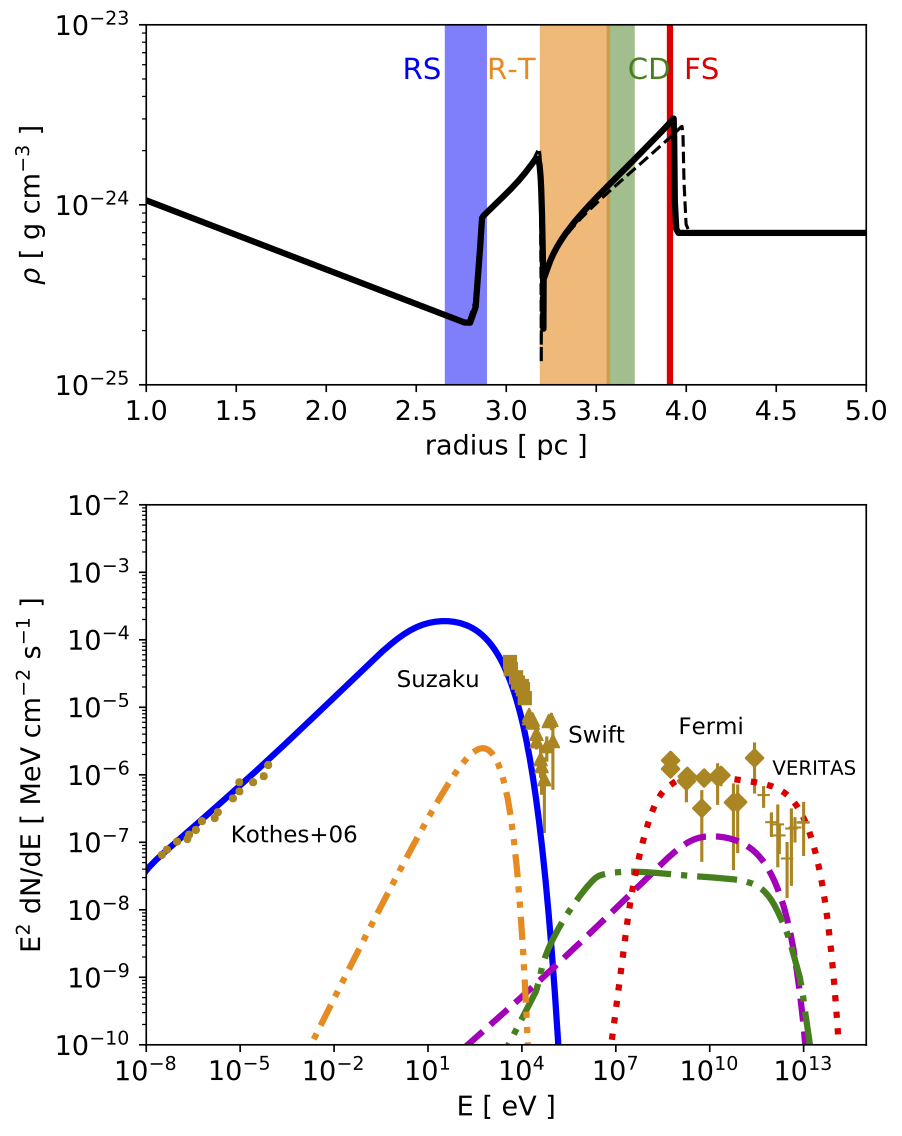}
\caption{
Upper panel: the solid line shows the calculated density distribution from Model A0 at an age of 446 yr. The color bands show the observed ranges of FS (red), CD (green) and RS (blue) radii of Tycho's SNR taken from \cite{2005ApJ...634..376W}. The orange band indicates the extent of expected R-T mixing \citep{2001ApJ...549.1119W}. The dashed line is the results from an identical model but without including CR feedback. Lower panel: the corresponding calculated non-thermal SED decomposed into its individual emission components, including synchrotron (blue solid), thermal bremsstrahlung (orange dash-dot-dotted), non-thermal bremsstrahlung (green dash-dotted), IC (magenta dashed) and $\pi^0$-decay (red dotted). The data points show the currently available observed fluxes - radio observations \citep[][dotted]{2006AA...457.1081K}, X-ray observations by {\it Suzaku} \citep[][and references therein, square]{2012ApJ...744L...2G} and {\it Swift/BAT} \citep[][triangle]{2014ApJ...797L...6T}, and $\gamma$-ray observations by {\it Fermi}-LAT \citep[][rhombus]{2012ApJ...744L...2G, 2017ApJ...836...23A} and {\it VERITAS} \citep[][cross]{2011ApJ...730L..20A, 2017ApJ...836...23A}.
} 
\label{test:tycho}
\end{figure}

\begin{figure}
\centering
\includegraphics[width=8cm]{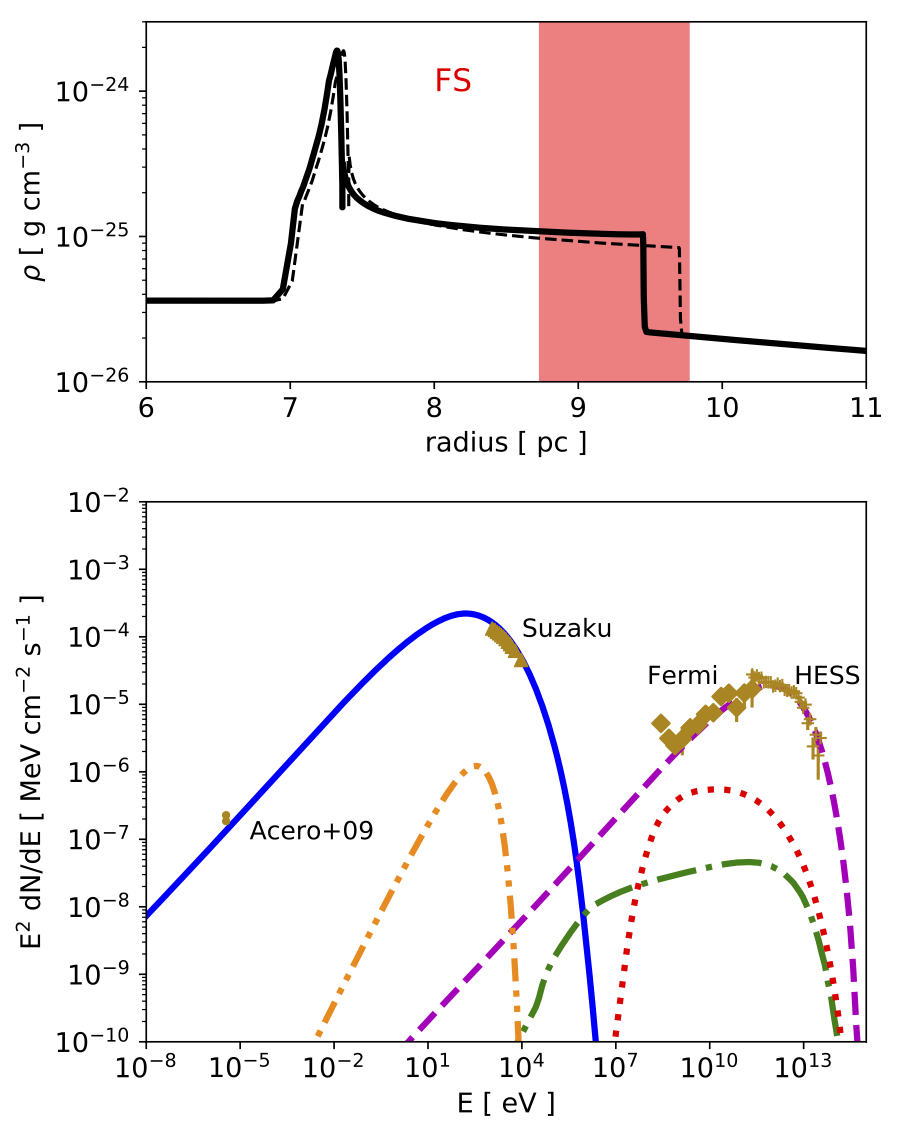}
\caption{
Results from Model B0. The format is the same as Figure~\ref{test:tycho}. A normalization factor $f_\mathrm{norm}=0.7$ has been applied to all calculated emission components to match the observed flux. Data points in lower panel:  radio observations \citep[][dotted]{2009AA...505..157A}, X-ray observation by {\it Suzaku} \citep[][triangle]{2004ApJ...602..271L} , $\gamma$-ray observation by {\it Fermi}-LAT \citep[][diamond]{2007AA...464..235A,2011AA...531C...1A,2018AA...612A...6H} and H.E.S.S. \citep[][cross]{2018AA...612A...6H}.
}
\label{test:RXJ1713}
\end{figure}

Tycho is identified to be the remnant produced by the historical supernova SN1572 which is classified as a Type Ia from its light-echo spectrum, chemical abundance pattern inferred from the X-ray spectrum and so on. Although it has been suggested that the ambient density around Tycho has an azimuthal gradient \citep{2013ApJ...770..129W}, we here assume a uniform ambient medium for simplicity. Fig.~\ref{test:tycho} shows the hydrodynamical and spectral results from our best-fit calibration model. The upper panel of Figure~\ref{test:tycho} shows the radial profile of the total mass density of the plasma (i.e., shocked/unshocked ISM and ejecta) at the current SNR age, $t_\mathrm{age}$ = 446 yr. The solid line is the result from Model A0, and the thin dashed line is the result when particle acceleration is not included but otherwise identical to Model A0. The red, blue and green bands are the radii of the forward shock (FS), reverse shock (RS), and contact discontinuity (CD) inferred from observation \citep{2005ApJ...634..376W}. We can see that our simulation can reproduce the FS and RS positions \footnote{As mentioned in Section 2, we only consider DSA at the FS in this work as a smoking-gun evidence of efficient DSA at the RS in SNRs is still absent. We can see in the upper panel of Fig. 2 the difference between the solid line and the dashed line which shows the results with and without feedbacks from an efficient DSA at the FS, respectively. DSA at the RS can be included in the code relatively easily when such evidence will surface in the future. }, but not the case for the CD. If particle acceleration is efficient (i.e. small $\gamma_\mathrm{eff}$), however, it has been reported that Rayleigh-Taylor (R-T) instability can develop between the FS and CD \citep[e.g.,][]{2001ApJ...560..244B,2013MNRAS.429.3099W} and the CD position can possibly extend outward significantly \citep[also see discussions in][]{Slane2014}. Our calculation can hence be considered to be in good agreement with observations on dynamics. The calculated SED at the same age is plotted in the lower panel. The observed fluxes are overlaid in the same plot. The agreement is found to be reasonable and reproduces the result of \citet{Slane2014} in their Model A. It can be seen that Tycho has a soft GeV-to-TeV spectrum from {\it Fermi} and {\it VERITAS} data, which can be explained by a $\pi^0$-decay origin with a softer than $E^{-2}$ underlying proton spectrum, but the mechanism of spectral softening of the accelerated protons relative to the canonical $E^{-2}$ prediction of DSA at a strong shock is not yet well understood.

RX~J1713 is believed to be the product of SN393 which has been classified as a CCSN, and the SNR has been well detected in multi-wavelength observations. The origin of the bright $\gamma$-ray emission from RX J1713 is still being intensively discussed as mentioned above. \cite{2012ApJ...746...82F} reported that the azimuthal distributions of H I and H$_2$ gases are consistent with the morphology of the observed TeV $\gamma$-rays, suggesting a hadronic origin. The gas distribution exhibits a low-density cavity surrounded by a dense shell, which has been suggested to be the result of the stellar wind of the progenitor prior to SN explosion inside a dense gas cloud. One the other hand, the observed hard $\gamma$-ray spectrum and the absence of optical signatures of the shock interacting with dense gas support a leptonic origin. In Model B0, we adopt a simple power-law $\rho \propto r^{-2}$ CSM model without considering the possibility of shock-cloud interaction, which is similar to the best-fit model for RX J1713 presented in \citet{LEN2012}. The results are summarized in Fig.~\ref{test:RXJ1713}, which shows the time snapshots of gas density profile and emission SED at $t_\mathrm{age}$ = 1625 yr. The FS position observed by {\it Fermi}-LAT \citep{2016ApJS..224....8A} is shown by the red shaded region in the upper panel, which is consistent with the model. The CD and RS locations for this remnant are not well constrained due to the very faint X-ray emission from the ejecta. The radio and non-thermal X-ray spectra can be well-reproduced by the model, and the hard observed $\gamma$-ray spectrum is well-reproduced by an IC origin. The results are found to be consistent with the model by \citet{LEN2012}.

Our result that the $\gamma$-ray emission is dominated by the IC component can be understood by considering the spatial distribution of the ambient gas density and magnetic field. The global magnetic field $B_0$ is as low as $\sim\ 6.6\times 10^{-2} (\sigma_\mathrm{w}/0.004)^{1/2}(\dot{M}/7.5\times10^{-6}\ M_\odot/\mathrm{yr})^{1/2}(V_\mathrm{w}/20\ \mathrm{km}/\mathrm{s})^{1/2}(r/9.5\ \mathrm{pc})^{-1}\ \mu \mathrm{G}$ at $t_\mathrm{age}$ = 1625 yr, so the amplified magnetic field $B_2$ is also moderate as $\sim\ 6.4\ \mu\mathrm{G}$ at the same time. This amplified but relatively low B-field behind the shock leads to an inefficient synchrotron loss such that the electrons can be accelerated to momenta capable of powering the observed $\gamma$-rays through the IC mechanism. Meanwhile, the ambient gas density also decreases rapidly as the SNR expands into the wind, so that the $\pi^0$ decay component is effectively suppressed.   

We also note that we applied a flux normalization factor, $f_\mathrm{norm}=0.7$, for all calculated emissions to match the observations mainly for two reasons; the distance of the SNR and a volume filling factor. We assume 1.0 kpc as the distance of RX~J1713 in this work, which involves uncertainty. While our models assume spherically symmetry, many SNRs like RX~J1713 are not a perfect spherical shell in gamma-rays \citep[see, e.g., Fig.1(a) in][]{2012ApJ...746...82F}. In other words, a prefactor $< 1$ has to be applied to our spectral SED to account for this volume filling factor. These uncertainties can be interpreted as the possible origins of $f_\mathrm{norm}$ in our models.

Based on our calibrated models A0 and B0 for an Ia and a CC SNR respectively, we now parametrically study the time-evolution of broadband non-thermal SED from SNRs interacting with different ambient environments. We note that we have chosen an ejecta mass of $M_\mathrm{ej} = 3.0\ M_\odot$ to calibrate with RX~J1713 in Model B0, but the ejecta mass can vary for different CC SN progenitors and SN types \citep[e.g.,][]{2015MNRAS.452.3869N}. Therefore, for Model B with a power-law CSM environment, we will survey over two ejecta models with $M_\mathrm{ej} = 3.0, 10.0\ M_\odot$ respectively. Other parameters are kept identical to Model A0 and B0 unless otherwise specified. The model parameters are summarized in Table~\ref{table:param}.

\subsection{Ia SNR models with a uniform ISM-like ambient medium} \label{sub:ISM}

\begin{figure*}
\centering
\includegraphics[width=18cm]{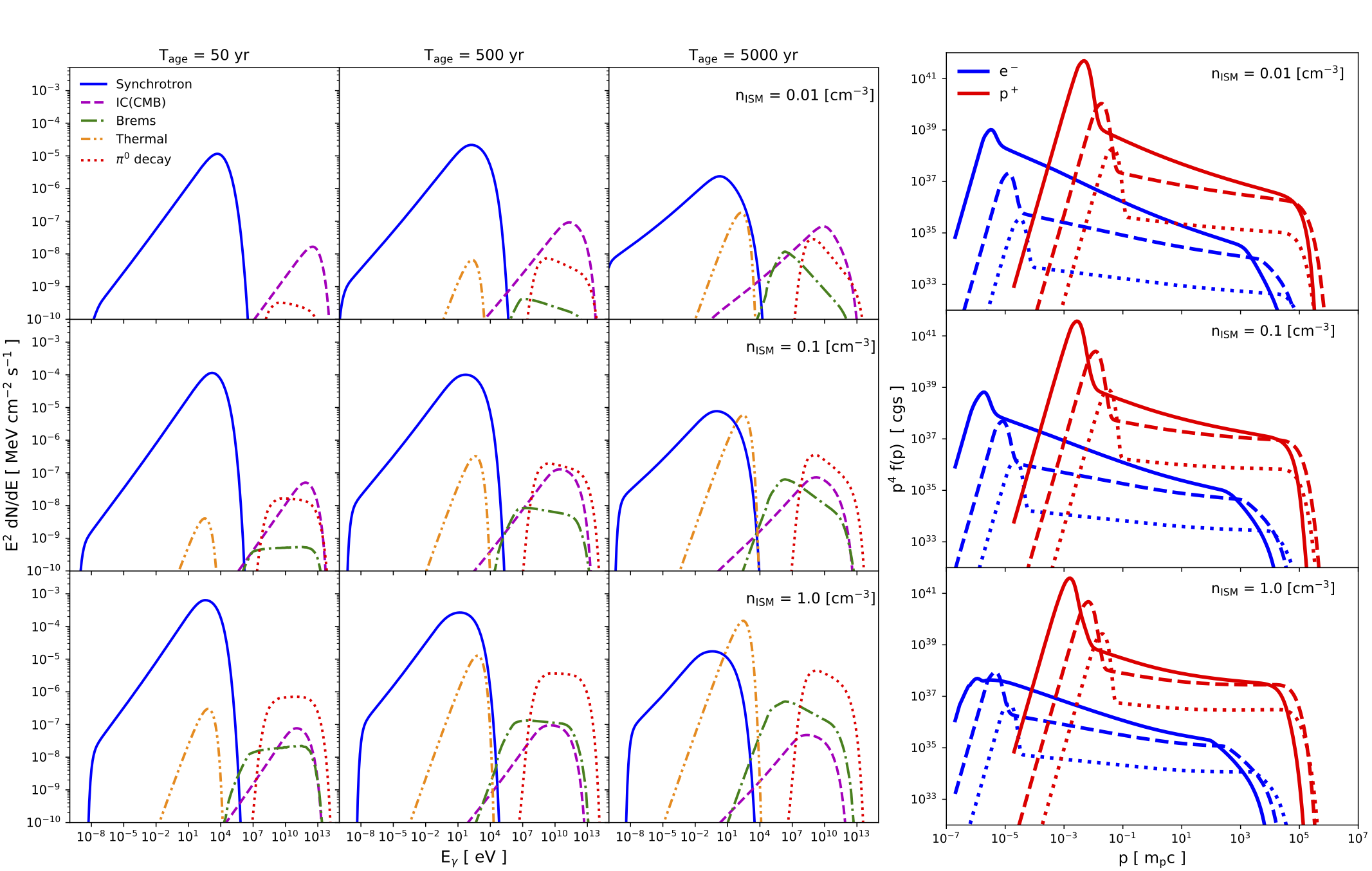}
\caption{
Left panel: time evolution of volume-integrated broadband SED from Type Ia SNR models with different ISM densities. Here $t_\mathrm{age}$ = 50, 500, 5000 yr moving from left to right panels, and $n_\mathrm{ISM}$ = 0.01, 0.1, 1.0 cm$^{-3}$ from top to bottom which correspond to Model A1, A2 and A3 respectively. The line formats are identical to the lower panel of Fig.~\ref{test:tycho} and Fig.~\ref{test:RXJ1713}.  
Right panel: time evolution of volume-integrated proton (red) and electron (blue) distribution functions with different ISM densities. Here $n_\mathrm{ISM}$ = 0.01, 0.1, 1.0 cm$^{-3}$ moving from top to bottom, and the dotted, dashed and solid lines correspond to $t_\mathrm{age}$ = 50, 500 and 5000 yr, respectively. 
} 
\label{multi:n}
\end{figure*}

\begin{figure}
\centering
\includegraphics[width=8.8cm]{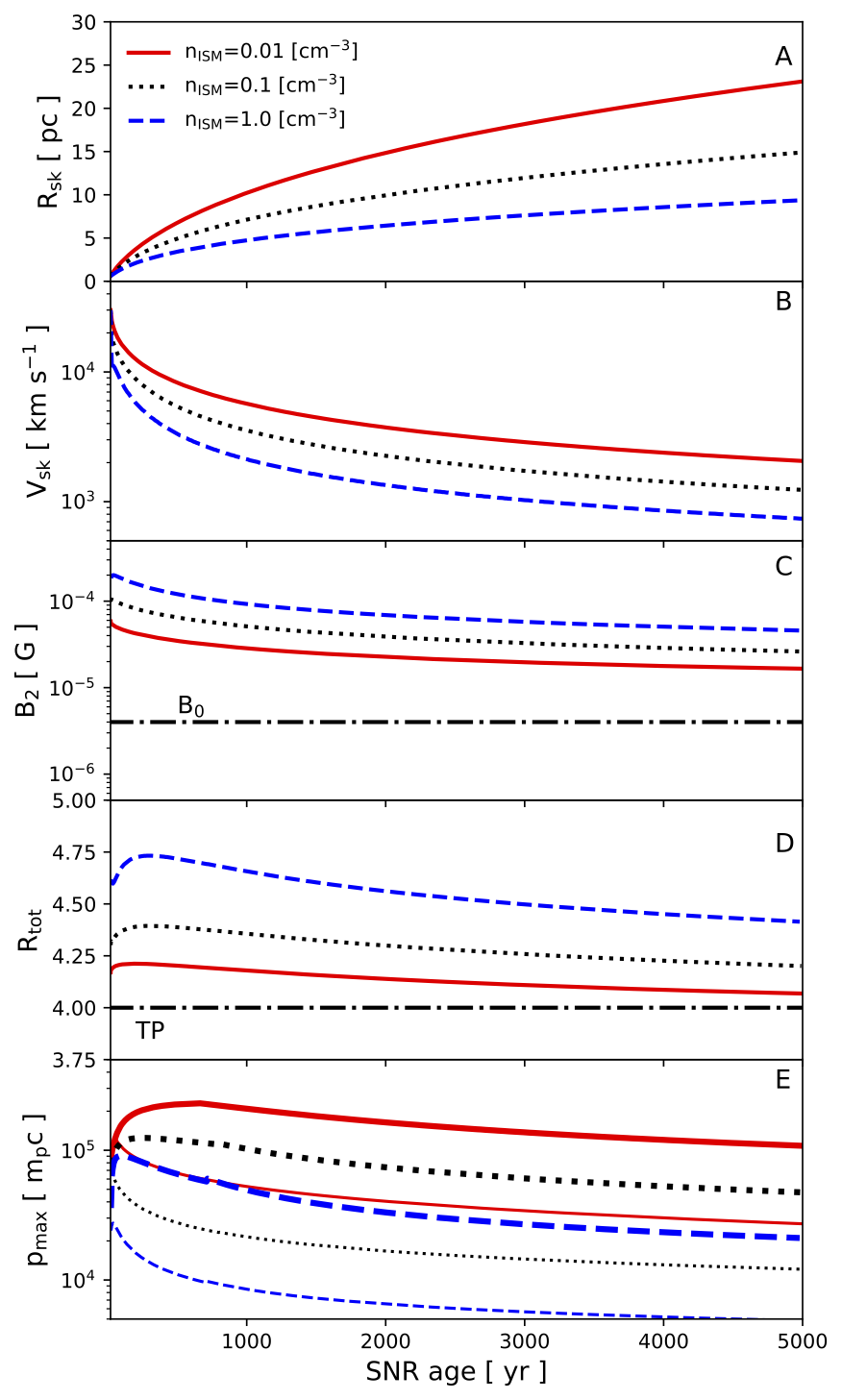}
\caption{
Time evolution of hydro and DSA outputs from our Type Ia models A1, A2 and A3 in three different ISM densities. Panel A shows the forward shock radius $R_\mathrm{sk}$, panel B shows the forward shock velocity $V_\mathrm{sk}$, panel C shows the magnetic field strength immediately downstream from the forward shock, panel D shows the total shock compression ratio, and panel E shows the maximum momentum $p_\mathrm{max}$ of accelerated protons (thick lines) and electron (thin lines). In all panels, the red solid, black dotted and blue dashed lines correspond to the cases of $n_\mathrm{ISM}=0.01,\ 0.1,\ 1.0\ \mathrm{cm}^{-3}$, respectively. The dash-dotted line shows the value of the ambient magnetic field $B_0$ = 4.0 $\mu$G in panel C and the expected compression ratio from a test-particle (TP) approximation $R_\mathrm{tot} = 4$ in panel D.
} 
\label{out:n}
\end{figure}
The left panel of Fig.~\ref{multi:n} shows the time-evolution of the broadband SED from our Type Ia SNR models A1, A2 and A3 for three different ISM densities, and the right panel shows the corresponding evolution of the underlying CR distribution functions. Fig.~\ref{out:n} shows the time-evolution of important hydrodynamical and DSA outputs.
In the GeV to TeV energy range, as time evolves, the flux of non-thermal bremsstrahlung (green dash-dotted line) and $\pi^0$-decay (red dotted line) are found to be increasing monotonically, but there is not much accompanied brightening in the IC component (magenta dashed line). This difference in the evolution is mainly caused by the energy loss of the accelerated particles. The intensities of non-thermal bremsstrahlung, $\pi^0$-decay, and IC are proportional to the fluxes of the accelerated particles multiplied by the number density of their respective interaction targets, i.e., ISM gas for non-thermal bremsstrahlung and $\pi^0$-decay, and CMB photons for IC. However, the high-energy flux of the accelerated electrons is highly suppressed by a fast energy loss due to synchrotron radiation. The synchrotron loss time-scale, $t_\mathrm{syn} = 3m_\mathrm{e}c^2/4c\sigma_\mathrm{T}U_\mathrm{B}\gamma \sim 130(E_\mathrm{e}/10\ \mathrm{TeV})^{-1}(B/100\ \mathrm{\mu G})^{-2}\ \mathrm{yr}$ is comparable to the SNR age with the post-shock magnetic field $\sim 100\ \mu\mathrm{G}$ being highly amplified in the shock precursor relative to the unshocked magnetic field $B_0=4.0\ \mu\mathrm{G}$ due to an efficient CR acceleration (see, panel C and E of Fig.~\ref{out:n}). On the other hand, although the proton spectrum also suffers from energy-loss from p-p inelastic scatterings, even in the denser case of $n_\mathrm{ISM}=1.0\ \mathrm{cm}^{-3}$, the energy loss time-scale $t_\mathrm{pp}=1/\sigma_\mathrm{pp}v_\mathrm{p}n_\mathrm{SNR}\sim 3\times10^7R_\mathrm{tot}^{-1}(n_\mathrm{ISM}/1\mathrm{cm}^3)^{-1}\ \mathrm{yr}$ is still much longer than the SNR age, so the effect is not significant on the protons. As a result, the peak of the IC spectrum shifts to lower energy in the early phase due to the fast synchrotron loss, and the peak flux does not vary much as the SNR ages. 

From the trend of flux evolution, we can see an interesting {\it leptonic}-to-{\it hadronic} transition in the moderately dense ISM case $n_\mathrm{ISM}$ = 0.1 cm$^{-3}$ at a few 100~yr. The middle panels in Fig.~\ref{multi:n} shows that the dominant flux of TeV range is IC at $t_\mathrm{age}$ = 50 yr, while $\pi^0$ flux becomes comparable to IC at $t_\mathrm{age}$ = 500 yr, and finally $\pi^0$ surpasses IC at $t_\mathrm{age}$ = 5000 yr. On the other hand, in the thin and dense ISM cases, the transition does not happen within a few 1000~yr. This behavior is mainly dictated by the gas density interacting with the shock (see the uppermost and lowermost panels in Fig.~\ref{multi:n}). 

We also see a systematic steepening of the $\gamma$-ray spectra with age in all models, which reflects the steeping of the proton spectrum from eq.~(\ref{eq:fp}). This effect comes from the deceleration of the FS with time due to an asymptote from free-expansion phase to Sedov phase. As a result, $v_A$ which is high due to the amplified $B$-field becomes non-negligible compared to the gas velocities in the later phase, and the effective compression ratio $S_{tot}$ is suppressed. The $\gamma$-ray spectrum hence becomes steeper with time because the spectral index of the particle distribution function is roughly proportional to $3S_\mathrm{tot}/(S_\mathrm{tot}-1)$ \citep{CBAV2009}. The steepening is even more prominent in the thin ISM case since $v_A \propto \rho^{-1/2}$ is larger in these models.

\subsection{CC SNR models with a power-law CSM-like ambient medium}\label{sub:CSM}
\subsubsection{3 M$_\odot$ case}

\begin{figure*}
\centering
\includegraphics[width=18cm]{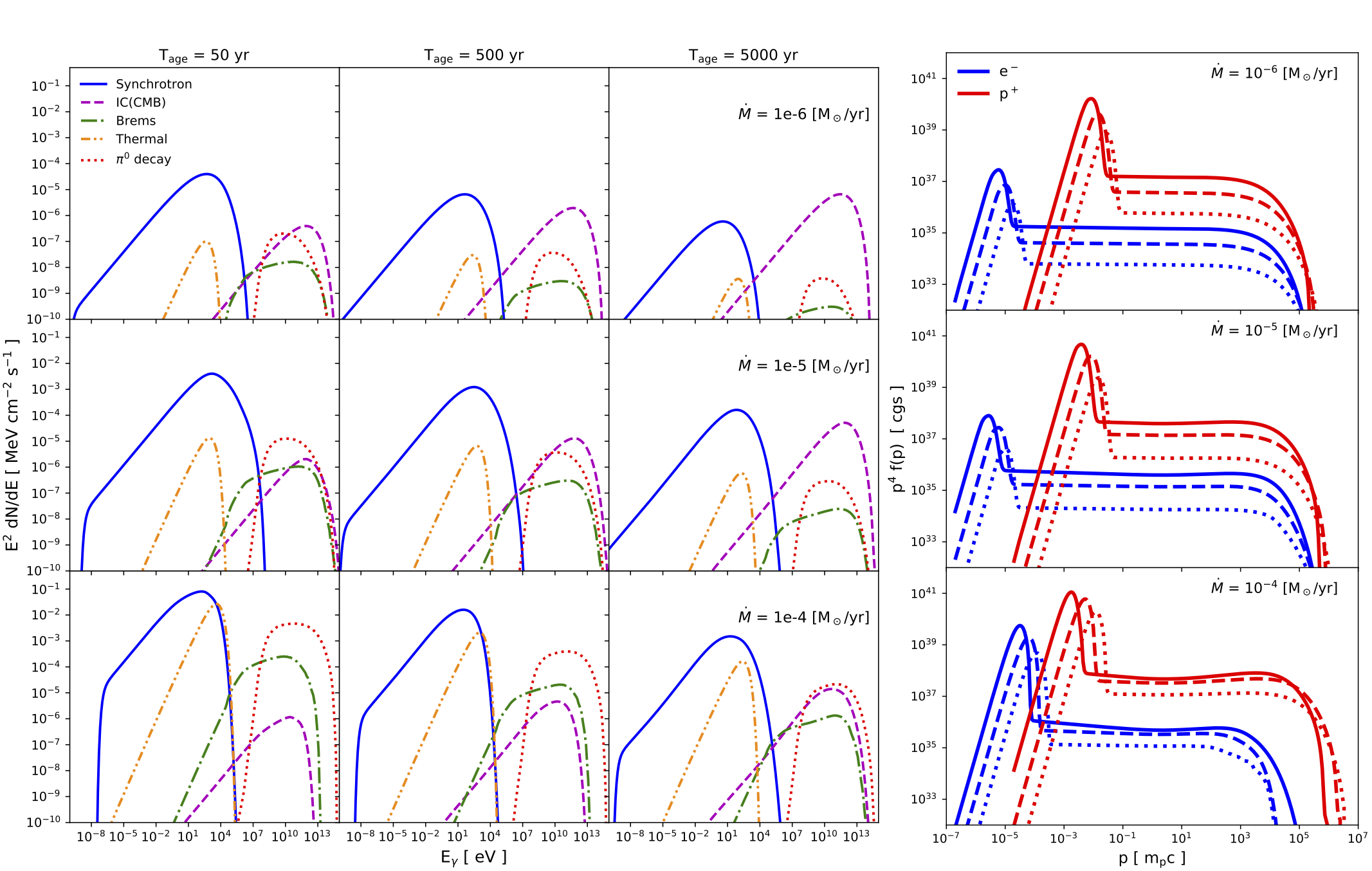}
\caption{
Left panel: SED evolution for our CC SNR models in different CSM environments. $t_\mathrm{age}$ = 50, 500, 5000 yr from left to right and $\dot{M} = 10^{-6},\ 10^{-5},\ 10^{-4}\ M_\odot/\mathrm{yr}$ from top to bottom which corresponds to Model B1, B2 and B3 respectively. The lines shown have the same format as in left panel of Fig.~\ref{multi:n}. Right panel: time-evolution of CR distribution function. $\dot{M} = 10^{-6},\ 10^{-5},\ 10^{-4}\ M_\odot/\mathrm{yr}$ from top to bottom. The lines shown have the same format as in right panel of  Fig.~\ref{multi:n}.
} 
\label{multi:m}
\end{figure*}
\begin{figure}
\centering
\includegraphics[width=8.8cm]{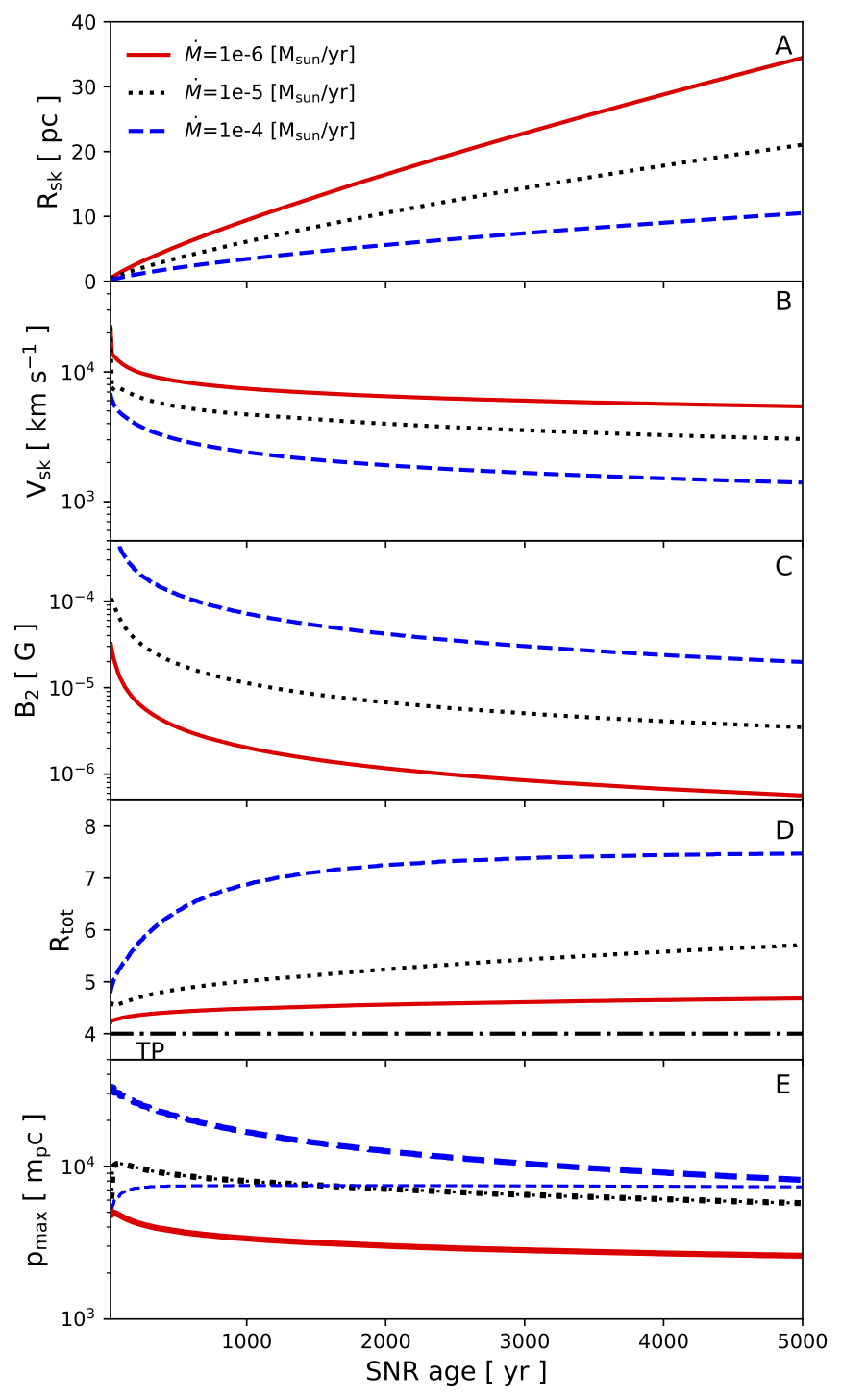}
\caption{
Time evolution of hydro and DSA outputs from model B1, B2, and B3. In all panels, the red solid, black dotted and blue dashed lines correspond to the cases of $\dot{M}=10^{-6}, 10^{-5}, 10^{-4}\ M_\odot/\mathrm{yr}$ respectively. The line formats are identical to Fig~\ref{out:n}.
} 
\label{out:m}
\end{figure}
Here we simulate the SNR evolution in a power-law CSM inside which a CCSN explodes with an ejecta mass of $M_\mathrm{ej} = 3.0\ M_\odot$. The results are shown in Fig.~\ref{multi:m} and Fig.~\ref{out:m}. The models correspond to a CSM with $\dot{M}$ = 10$^{-6}$, 10$^{-5}$, 10$^{-4}$ M$_\odot$/yr for Model B1, B2 and B3 respectively. In the GeV-TeV spectrum, the IC flux increases, while the $\pi^0$-decay and non-thermal bremsstrahlung fluxes decrease as time proceeds, which is an opposite behavior compared to the uniform ISM models we see earlier. Since the CSM has a power-law density distribution in these CC SNR models, the CSM provides a dense target for producing $\pi^0$-decay and bremsstrahlung photons effectively which dominate the spectrum in the GeV-TeV range in the very early phase after the explosion.  However, as the CSM density decreases as $r^{-2}$, as time passes and the shock propagates through the wind material, the target gas density becomes low quickly, so the emission efficiency through $\pi^0$-decay and bremsstrahlung is suppressed accordingly. Moreover, the accelerated particles advected downstream from the shock also suffer from adiabatic loss to a larger extent than Model A due to the fast expansion of the SNR in a $\rho \propto r^{-2}$ wind. As a result, the fluxes of $\pi^0$-decay and non-thermal bremsstrahlung constantly decrease with age. On the contrary, the target photons of IC which is CMB here is homogeneous in space. The magnetic field is also lower than those we see in the uniform ISM cases (see panel C of Fig.~\ref{out:n} compared to that of Fig.~\ref{out:m}). This means synchrotron loss is less important in the CC SNR models. Indeed, the quick shift of the peak energy of the synchrotron and IC components which is seen in Fig.~\ref{multi:n} does not occur here, and the IC emission gradually increases with time with the peak staying at more-or-less the same energy range. These are the main reasons why the IC photons are constantly produced in the power-law CSM cases. In the center and bottom panels of Fig.~\ref{multi:m}, we can also see a low-energy cutoff in the synchrotron spectrum in the radio band. This is because free-free absorption is efficient at early time due to the dense unshocked CSM in front of the shock. The absorption becomes inefficient with time however as the SNR shock propagates into the relatively thin region of the CSM, so the cutoff shifts to lower frequencies. 

Our SED evolution model for the CCSN with power-law CSM cases suggests a {\it hadronic}-to-{\it leptonic} transition in the GeV-TeV range if the wind density is moderately dense with a $\dot{M}$ = 10$^{-5}$ M$_\odot$/yr, which is again the exact opposite behavior we see in the Ia SN cases with uniform ISM. We suggest that these contrasting spectral evolution and transition of dominant $\gamma$-ray component can be useful for probing the surrounding environment of SNRs, especially in the near future as the sample of $\gamma$-ray SNR observations is enlarged by future observatories such as the upcoming observatories such as CTA (see section 3.6).

\subsubsection{10 M$_\odot$ case}

\begin{figure}
\centering
\includegraphics[width=8.8cm]{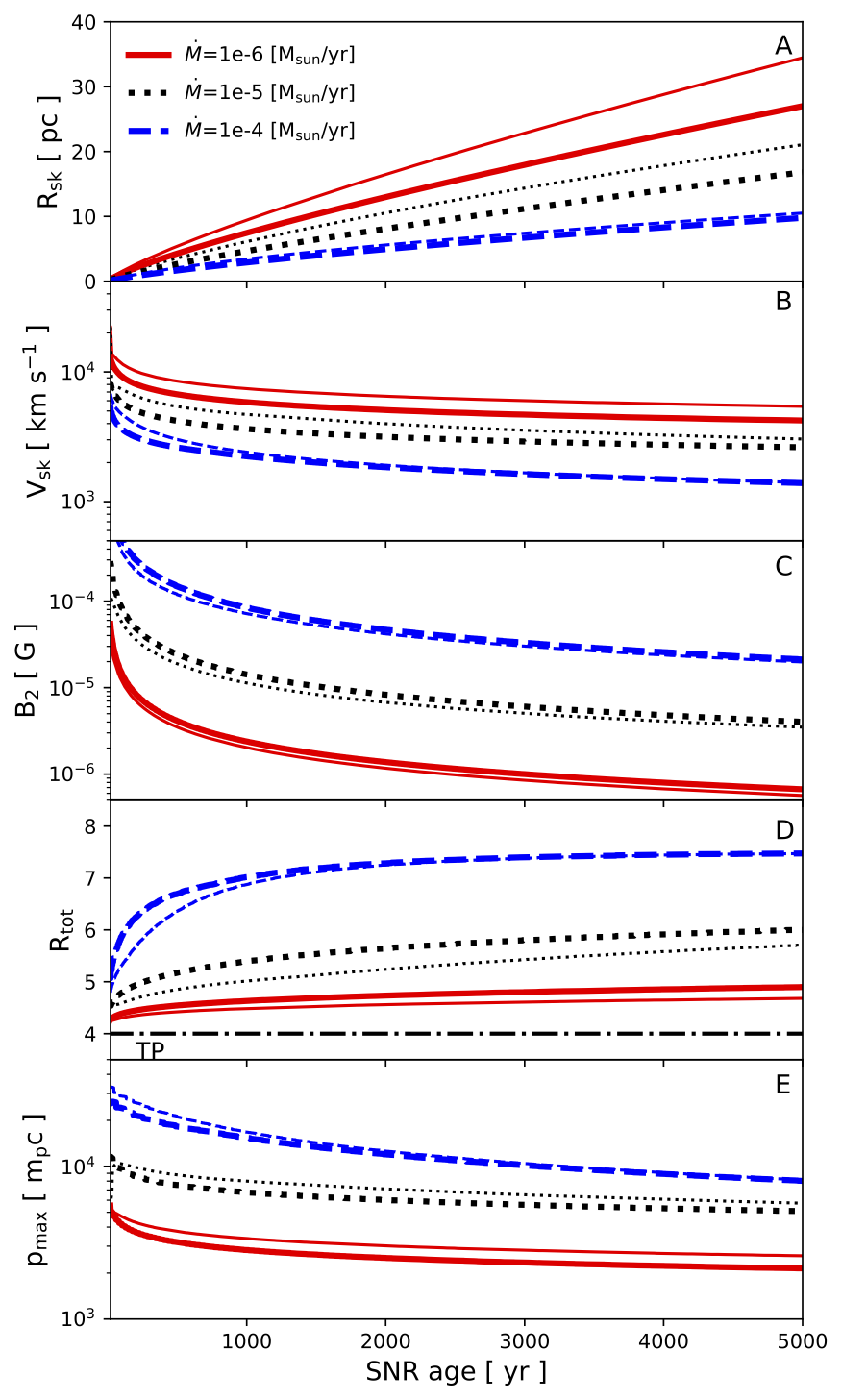}
\caption{
Same as Fig.~\ref{out:m}, but with an ejecta mass of $M_\mathrm{ej}=10.0\ M_\odot$. Only the $p_\mathrm{max}$ of protons is plotted here in panel E for clarity. The thick red solid, black dotted, and blue dashed lines show the results of model B4, B5, and B6. For comparison, thin lines show the results of the $3M_\odot$ case as in Fig.~\ref{out:m}.} 
\label{out:10m}
\end{figure}
The results from our CCSN models with an ejecta mass $M_\mathrm{ej}=10\ M_\odot$ are shown in Fig.~\ref{out:10m}. For comparison, the results of $M_\mathrm{ej}=3\ M_\odot$ are also overlaid. From panel A and B, it can be seen that while the shock dynamics for the case of $\dot{M}$ = 10$^{-6}$ M$_\odot$/yr and $\dot{M}$ = 10$^{-5}$ M$_\odot$/yr are affected by a different ejecta mass, the results for $\dot{M}$ = 10$^{-4}$ M$_\odot$/yr are nearly identical. These differences can be explained by a different evolutionary phase of the SNR at a given age. In the cases of $\dot{M}$ = 10$^{-6}$ M$_\odot$/yr and $\dot{M}$ = 10$^{-5}$ M$_\odot$/yr, the CSM density is relatively low and the mass swept up by the FS is smaller than the ejecta mass, the dynamics of these two cases thus follow the self-similar solution, $R_\mathrm{sk}\propto (E_\mathrm{SN}/M_\mathrm{ej}A)^{1/5}t^{4/5}$ and $V_\mathrm{sk}\propto(E_\mathrm{SN}/M_\mathrm{ej}A)^{1/5}t^{-1/5}$ \citep{1982ApJ...259L..85C,1982ApJ...258..790C}, where $A\equiv\dot{M}/4\pi V_w$, which depends on the ejecta mass. On the contrary, the CSM material in the case of $\dot{M}$ = 10$^{-4}$ M$_\odot$/yr are dense enough and the swept mass become comparable to the ejecta mass at $t_\mathrm{age}\le1000\ \mathrm{yr}$, the dynamics hence follows the Sedov solution, $R_\mathrm{sk}\propto (E_\mathrm{SN}/A)^{1/3}t^{2/3}$ and $V_\mathrm{sk}\propto(E_\mathrm{SN}/A)^{1/3}t^{-1/3}$ \citep{1959sdmm.book.....S}, which are independent of the ejecta mass.

As for the other quantities shown in Fig.~\ref{out:10m} like $B$-field and $p_\mathrm{max}$, the differences are found to be subtle only \footnote{The fact that the shock velocities in the 10 M$_\odot$ cases are lower than those of the 3 M$_\odot$ models at any given age implies that the effects of CR back-reaction and shock modification becomes important at an earlier phase. This is evident from the slightly higher total compression ratio and amplified magnetic field, as shown in panel C and D in Fig.~\ref{out:10m}. }. As a result, we do not see any remarkable difference in the non-thermal SED between the 3 M$_\odot$ and 10 M$_\odot$ models. We can conclude that it is hard to distinguish the progenitor from the non-thermal emission in the SNR phase, and other information which reflect the progenitor properties such as thermal X-ray emission lines are needed to link an observed SNR to its progenitor origin. In section~\ref{comparison} where we compare our results to observations, we will only show the results from the 3 M$_\odot$ models because of this insensitivity of the $\gamma$-ray emission to ejecta mass.

\subsection{A case of pre-SN enhanced mass loss}\label{sub:en}

\begin{figure*} 
\centering
\includegraphics[width=15cm]{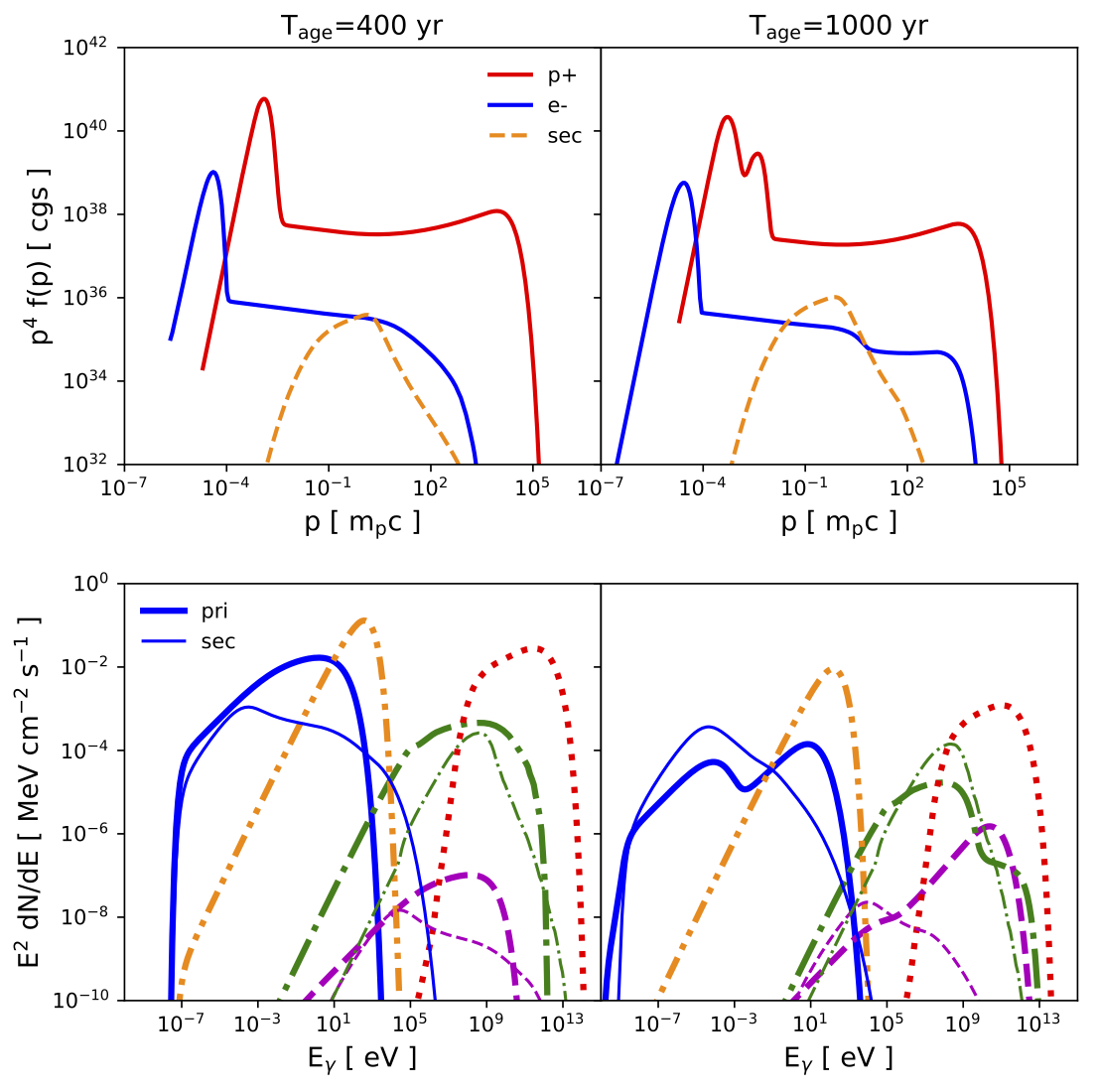}
\caption{
Time-evolution of the volume-integrated distribution function of the accelerated particles (upper panels) and broadband emission SED (lower panels), which correspond to Model C. Left and right panels show the results at $t_\mathrm{age}$ =400, 1000 yr, respectively. In the upper panels, the red solid, blue solid and orange dashed lines correspond to the primary protons, electrons, and secondary electrons/positrons, respectively. In the lower panels, the thick lines show the emission produced by the primary particles, and the thin lines are by the secondaries. The line colors are the same as Fig.~\ref{multi:n}.
} 
\label{multi:e2}
\end{figure*}
\begin{figure}
\centering
\includegraphics[width=8.8cm]{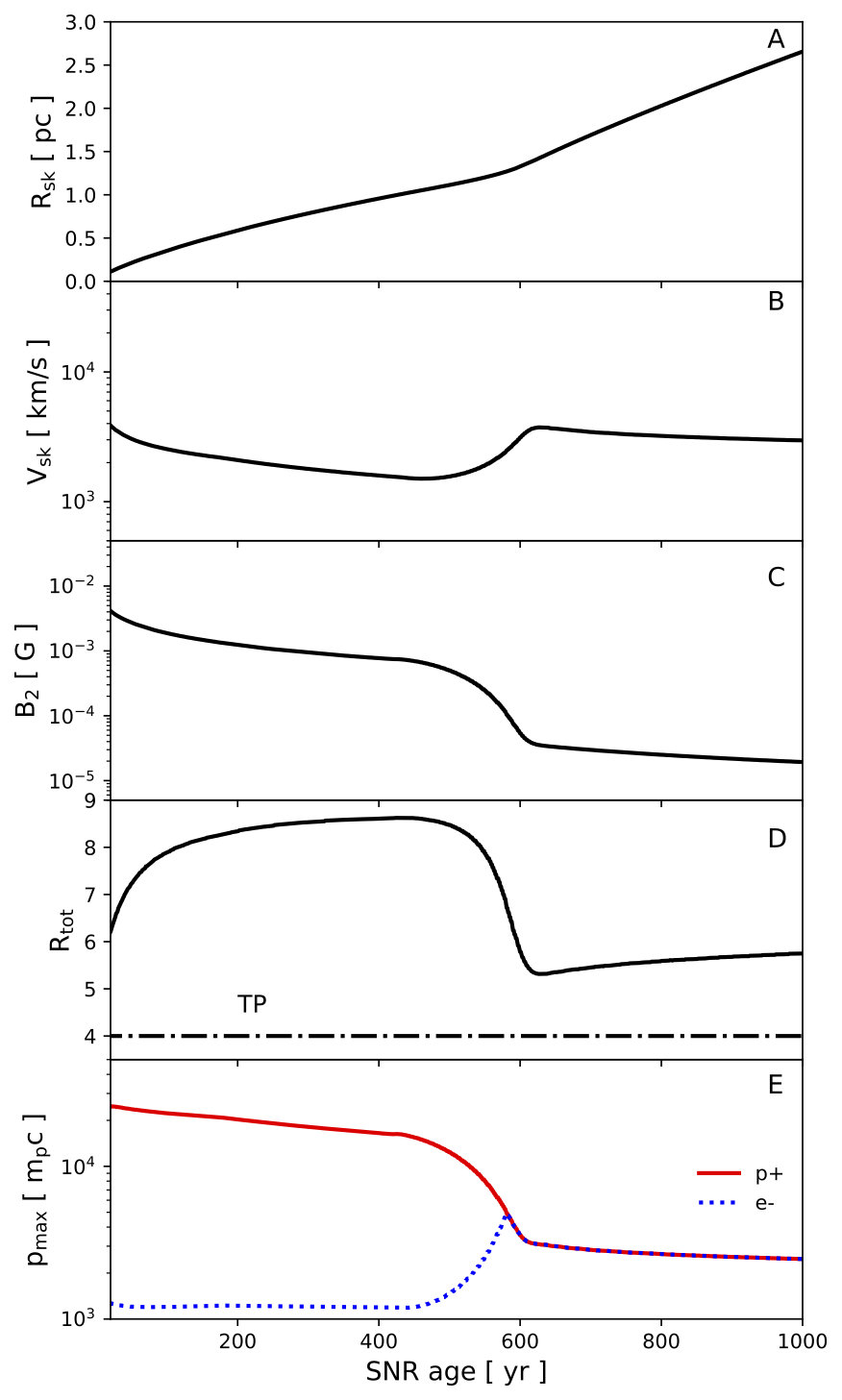}
\caption{
Time evolution of various values using model C. The values of each panel shows are the same as Fig.~\ref{out:n} and Fig.~\ref{out:m}.
} 
\label{out:e2}
\end{figure}
Results from Model C where the $\dot{M}$ is boosted to $10^{-2}$ M$_\odot$/yr in the last 1000~yr before CC are shown in Fig.~\ref{multi:e2} and Fig.~\ref{out:e2}. Fig.~\ref{multi:e2} shows the phase-space distributions of the accelerated primary (solid) and secondary (dashed) particles in the upper panels, and the photon SEDs in the lower panels. We choose to show the results at $t_\mathrm{age}$ =  400 (left panels) and 1000 (right panels) yr because these time epochs represent the phases before and after the shock has crossed the interface at $R_\mathrm{tr}$ between the dense confined CSM and the less dense wind outside. In this model, the shock reaches $R_\mathrm{tr}$ at $t_\mathrm{age} \sim$ 600 yr (see, Fig.~\ref{out:e2}). 

As the FS propagates in the region of dense CSM material with density $n \sim$ 10$^4$ cm$^{-3}$ and a high magnetic field $\ge$ 1 mG (see, panel C of Fig.~\ref{out:e2}), the electron maximum momentum $p_\mathrm{max, e}$ is determined by the energy loss time-scale $t_\mathrm{loss}$ rather than the age or escape time-scale because $t_\mathrm{loss} \sim t_\mathrm{syn} \sim 12 B_{-3}^{-2} E_{12}^{-1}$ yr, where  $B_{-3} = B_2 / 10^{-3}\mathrm{G}$ and $E_{12} = E_e / 10^{12}\mathrm{eV}$, is less than the SNR age at a given time (see, panel E of Fig.~\ref{out:e2}). However, as the FS breaks out from the dense inner shell, $p_\mathrm{max, e}$ is now limited by their escape through the FEB because the shock velocity is now restored to $\sim$ 4000 km/s and the magnetic field decreases to $\sim$ 10 $\mu$G (see, panel B, C and E in Fig.~\ref{out:e2}). Therefore, we can see two cutoffs at $p \sim 10\ m_\mathrm{p}c$ and $p \sim 10^3\ m_\mathrm{p}c$ in the volume-integrated electron spectrum (upper right panel in Fig.~\ref{multi:e2}) while the proton spectrum has one cutoff only at $p \sim 10^4 m_\mathrm{p}c$. These effects of a transition from a dense wind to a lower density wind also reflects in the spectra of synchrotron, non-thermal bremsstrahlung and IC emission (see, lower right panel in Fig.~\ref{multi:e2}). 

In the radio range of the SED, before the shock breaks out from the dense region, the dominant component is synchrotron radiation from the primary electrons (solid) and a spectral cutoff can be seen at $E_\gamma \sim10^{-7}\ \mathrm{eV}$ due to a strong free-free absorption. However, after the breakout, the dominant component is now the synchrotron emission from the secondary electrons and positrons. The reason is as follows. Electrons accelerated earlier on in the dense wind suffer from rapid energy loss through synchrotron emission and adiabatic expansion, and the freshly accelerated electrons in the outer tenuous wind have a higher $p_\mathrm{max}$ as mentioned above, but the synchrotron radiation from these freshly accelerated electrons is relatively weak due to a lower magnetic field in the tenuous wind, therefore the synchrotron flux from the primaries decreases with time. On the other hand, the contribution from the secondaries do not decrease as rapidly because these secondary particles are produced via $\pi^0$-decay not only by the freshly accelerated protons, but also by the proton accelerated earlier on in the dense wind continuously as the protons do not lose their energy as quickly as the electrons. This is why the transition from primary to secondary dominance happens in the synchrotron radiation.  

We suggest that this transition can potentially constrain the mass-loss history of massive stars. For example, the spectral index of synchrotron emission produced by the secondary particles is expected to be different from that produced by the primary electrons, which is evident from their very different distribution functions as shown in the upper panels of Fig.~\ref{multi:e2}. In particular, the synchrotron spectrum from the secondaries tends to be harder in the radio band.  In fact, hard radio indices are usually observed in older SNRs interacting with dense molecular clouds, such as IC443 \citep[e.g.,][]{2007AA...471..537C,2011AA...534A..21C}. These remnants are also believed to be producing a significant amount of secondaries. If a harder-than-expected spectral index will be observed in young SNRs which is not colliding with any dense cloud at the moment, it is possible that the SNR has evolved inside a dense confined CSM in the past, which can provide a hint on an enhanced mass loss of the progenitor star prior to CC. 

\subsection{Model versus data} \label{comparison}
We now try to compare our simulation results to observation results so far in terms of dynamics (e.g., shock radius and velocity) and $\gamma$-ray luminosity to check if our models are able to reproduce the bulk properties of observed SNRs. We include data like SN type, distance, shock radius, shock velocity, and radio, GeV, and TeV fluxes of SNRs from a younger age ($\sim100\ \mathrm{yr}$) to middle age ($\sim10,000\ \mathrm{yr}$). We summarize these data in Table.~\ref{table:data}. 
The data on SNR radii with errors are taken from the {\it Fermi} catalog \citep{2016ApJS..224....8A}, and those without errors are determined by the size of the radio remnants and are taken from the {\it SNRcat} \citep{2012AdSpR..49.1313F}.
The flux data is obtained again mainly from the {\it Fermi} and H.E.S.S \citep{2018AA...612A...6H} catalogs (see Table.~\ref{table:data} for details); those with errors are for detected SNRs, and those without errors are the upper limits of non-detected SNRs. 

\begin{figure*}
\centering
\includegraphics[width=18cm]{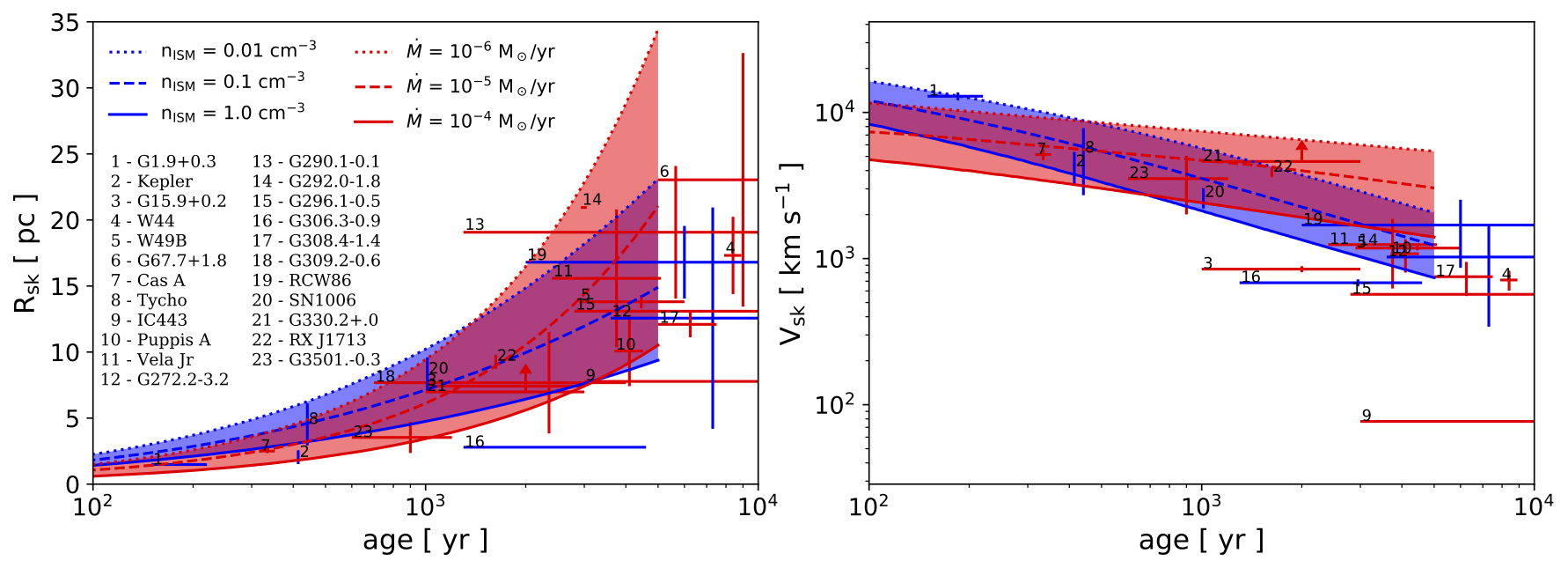}
\caption{
Left panel: FS location as a function of age. Blue (red) dotted, dashed, and solid line show the results of Model A1 (B1), A2 (B2), and A3 (B3), respectively. Right panel: FS velocity as a function of time.
The observation data are shown by the blue (Ia) and red (CC) data points in both panels, and are summarized in Table.~\ref{table:data}. 
} 
\label{t_vs_RV}
\end{figure*}
Fig.~\ref{t_vs_RV} shows the comparison of dynamical properties including the shock radius and shock velocity as a function of time from our models with observations. The blue data points are for the Type Ia SNRs and red ones are for the CC SNRs. We label each SNR by a number as summarized in left panel of Fig.~\ref{t_vs_RV}. In general, the overall trend of the observed distribution of shock radius and velocity as a function of SNR age can be explained by our simulation results for the parameter space we explored. There exist a few ``outliers'' which have small radii and velocities, which can be interpreted as SNRs interacting with a medium denser than what our models have considered. In fact, many of these are known to be interacting with dense molecular clouds at the moment. 

\begin{figure*}
\centering
\includegraphics[width=18cm]{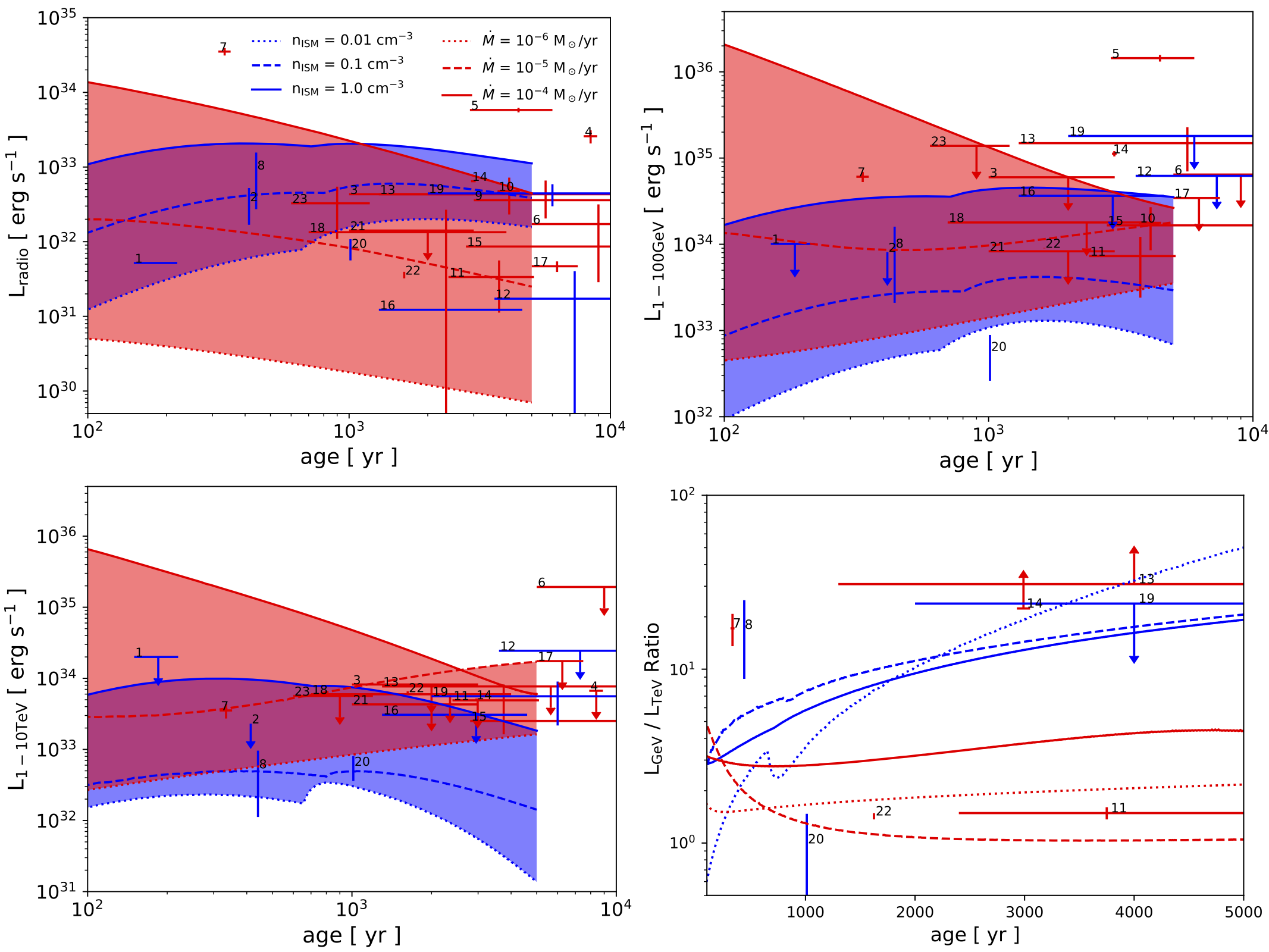}
\caption{
Upper left panel: radio luminosity at 1 GHz as a function of time. Upper right panel: integrated $\gamma$-ray luminosity from 1 GeV to 100 GeV as a function of time. Lower left panel: integrated $\gamma$-ray luminosity from 1 TeV to 10 TeV as a function of time. Lower right panel: the ratio of GeV to TeV luminosity as a function of time. The format of lines and data in all panels is the same as in Fig.~\ref{t_vs_RV}.
} 
\label{t_vs_L}
\end{figure*}
Fig.~\ref{t_vs_L} shows the time-evolution of the luminosity in three different energy ranges. The upper left panel shows the luminosity of 1 GHz radio continuum emission, which reflects the time-evolution of synchrotron spectrum for both the uniform ISM cases and power-law CSM cases (see also, Fig~\ref{multi:n} and Fig~\ref{multi:m}). Since synchrotron emissivity is proportional to the flux of accelerated electrons and the square of the local magnetic field strength, for the magnetic field which is constantly distributed in the uniform ISM cases, the synchrotron emissivity does not vary much and the spectral peak shifts to a lower energy with time, similar to the IC component previously discussed in sec.~\ref{sub:ISM}. On the other hand, in a power-law CSM case, the magnetic field decreases in proportion to $r^{-1}$, the synchrotron flux then also decreases with time just as the $\pi^0$-decay $\gamma$-rays do. Thus, radio luminosity increases gradually with time as a volume effect in the uniform ISM cases, but decreases in the power-law CSM cases. 

The upper right and lower left panels show the GeV luminosity integrated from 1 GeV to 100 GeV and TeV luminosity from 1 TeV to 10 TeV as a function of age, respectively. For the uniform ISM cases, while the GeV luminosity increases with time, TeV luminosity decreases. At the bottom end of the predicted flux which corresponds to the case of $n_\mathrm{ISM} = 0.01$~cm$^{-3}$, the $\gamma$-rays are dominated by IC at all time, and the decrease of the TeV flux can be understood as the energy loss of the highest-energy electrons. For the other two cases with a denser ISM, the trend reflects the time evolution of not only the normalization but also the shape of the $\pi^0$-decay spectra predicted by these models. As seen in Fig.~\ref{multi:n}, the $\pi^0$-decay spectrum become softer as time passes by. The reason has been discussed in the end of sec.~\ref{sub:ISM}, which is mainly because of the increasing importance of the effect from $v_A$, i.e., the velocity of the magnetic scattering centers. As the shock sweeps up more material as the SNR ages, the $\pi^0$-decay flux increases with time in general, but the TeV flux decreases due to a spectral softening of the underlying proton distribution. 

On the other hand, in the power-law CSM cases, both GeV and TeV luminosity decrease only in the case of the densest wind with $\dot{M} = 10^{-4} M_\odot / \mathrm{yr}$, but increase in the other two cases. This can be easily understood according to the discussion above in sec.~\ref{multi:m} on the evolution of IC and $\pi^0$-decay fluxes, and the dominant component is $\pi^0$-decay in the case of $\dot{M} = 10^{-4} M_\odot / \mathrm{yr}$. 

Here, to obtain the data points from the $\gamma$-ray observations, we assume that the observed spectra have a simple power-law distribution, so that the integrated luminosity can be calculated using the following expression, 
\begin{equation}
L_\Delta = 4\pi d_\mathrm{SNR}^2 \frac{(-\Gamma+1)(E_\mathrm{max}^{-\Gamma+2}-E_\mathrm{min}^{-\Gamma+2})}{(-\Gamma+2)(E_\mathrm{max}^{-\Gamma+1}-E_\mathrm{min}^{-\Gamma+1})}F_\Delta,
\end{equation}
where $F_\Delta, \Gamma, d_\mathrm{SNR}, E_\mathrm{min}, E_\mathrm{max}$ are the integrated flux, photon index, distance to the SNR, minimum and maximum energies of the integrated energy range. 

The radio luminosities from observations of both Type Ia and CC SNRs can be bulkly reproduced by our models with a few outliers such as Cas A and middle-aged SNRs interacting with MCs. 
As for the GeV and TeV observations, the statistics is still clearly very poor due to the small sample size of detected sources, so at the moment the comparison with the models is only preliminary. 
For older CC SNRs, a few outliers are found with
significantly higher luminosities than our results. These are again mostly middle-aged SNRs interacting with dense MCs which are not covered by our parameter space. 

The lower right panel shows the ratio of GeV to TeV luminosities which roughly quantifies the $\gamma$-ray spectral shape. Two trends can be seen in the result: one trend rises with time, and the other is nearly flat. These can be possibly explained by our discussion on flux evolution above. If $\pi^0$-decay is the dominant emission in $\gamma$-rays, the GeV luminosity increases and TeV luminosity decreases with time in uniform-ISM cases, and as a result, the ratio increases with time, while the ratio in power-law CSM case becomes nearly flat regardless of time because both GeV and TeV luminosities decrease with time, and  case. If, however, IC is the dominant contributor, the SED evolves without changing its shape, so the ratio does not vary in any significant way with time. Indeed, the observation data also appear to split into two regions; $L_\mathrm{GeV}/L_\mathrm{TeV} \sim 1$ and $L_\mathrm{GeV}/L_\mathrm{TeV} \ge 10$ despite the poor statistics. If both GeV and TeV emissions can be observed from an increased number of SNRs in the future, we will be able to see if the SNRs will segregate into two groups in this plot, which can make this quantity a useful probe of the ambient environment and hence the progenitor origin of SNRs.   

\subsection{Prospects for Cherenkov Telescope Array}

An instrument which can observe over a broad energy range from GeV to TeV energies with a high sensitivity, such as the Cherenkov Telescope Array (CTA), is ideal for a systematic investigation as introduced in this study. CTA can achieve an unprecedented sensitivity superior to existing detectors in the 20 GeV to 100 TeV energy range. With CTA, we expect that the number of detected $\gamma$-ray emitting SNRs will increase by roughly a factor of 10, which is essential for understanding the SNR population and their ambient environments. 

\begin{figure}
\centering
\includegraphics[width=8.8cm]{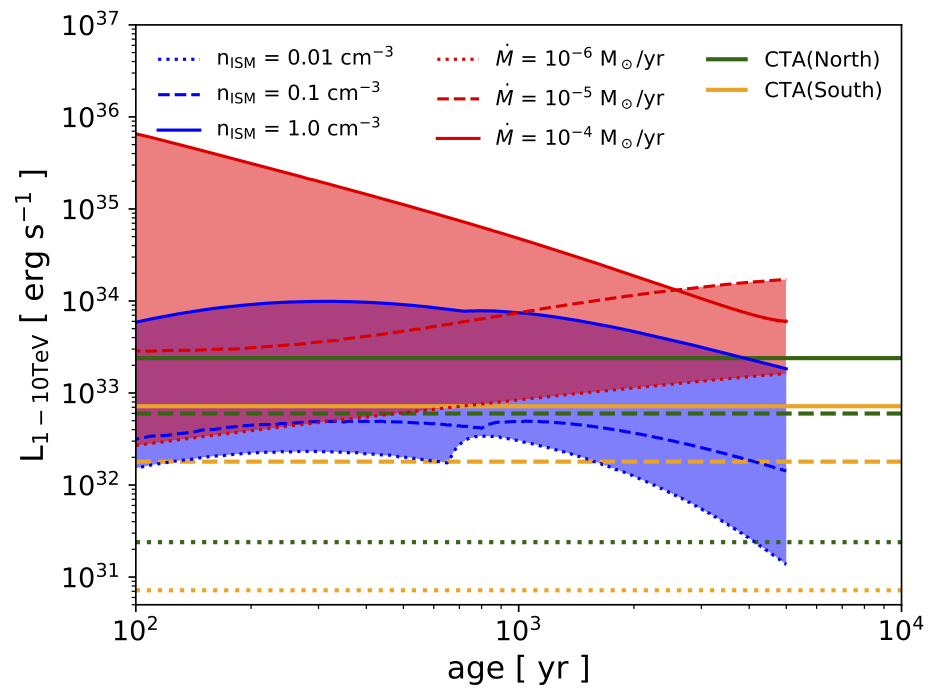}
\caption{
TeV luminosity as a function of time from our models compared with CTA sensitivities assuming different SNR distances. The green (yellow) lines show the sensitivities of CTA at the north (south) site for a source distance of 1.0 (dotted), 5.0 (dashed) and 10.0 (solid) kpc.}
\label{TeV_sens}
\end{figure}
Here, we compare the calculated TeV luminosity from our Ia and CC SNR models with the CTA sensitivities to predict the horizons for SNRs residing in different types of ambient environments. Fig.~\ref{TeV_sens} shows the range of model TeV luminosity and CTA sensitivities for different source distances. These sensitivities are calculated using the differential sensitivity curve assuming an observation time of 50 h 
(see,  for details, http://www.cta-observatory.org/science/cta-performance/ (version prod3b-v1)). We do not consider the possibility of source confusion, (fore-)background contamination and other complications for simplicity.

For SNRs with $d_\mathrm{SNR}=\ 1.0\ \mathrm{kpc}$, we see that, within the parameter range of our models, they are easily detectable regardless of age or ambient environment. For $d_\mathrm{SNR}=\ 5.0\ \mathrm{kpc}$, the detectability starts to depend on the SN type, age and environment. For both the southern and northern sky, the CC SNRs should be observable irrespective of age and environment. The Ia's are also detectable except for those in the southern sky with $t_\mathrm{age}\ge 2000\ \mathrm{yr}$ interacting with a very tenuous $n_\mathrm{ISM}\sim 0.01\ \mathrm{cm}^{-3}$ environment, or those in the northern sky with a density $n_\mathrm{ISM}\le\ 0.1\mathrm{cm}^{-3}$. 
For $d_\mathrm{SNR}=\ 10.0\ \mathrm{kpc}$, the sensitivity for the southern sky  is roughly the same as that of the northern sky for $d_\mathrm{SNR}=\ 5.0\ \mathrm{kpc}$. SNRs in the northern sky can be detected if the environment is dense, with $n_\mathrm{ISM}\sim 1.0\ \mathrm{cm}^{-3}$ for an Ia in a uniform ISM or $\dot{M}\ge10^{-5}\ M_\odot/\mathrm{yr}$ for a CC in a wind. So, let $n_\mathrm{ISM}=0.1\ \mathrm{cm}^{-3}$ and $\dot{M}=10^{-5}\ M_\odot/\mathrm{yr}$ to be the typical values for an ISM-like and a power-law CSM-like environments, respectively, we can conclude that the CTA has a sufficient sensitivity to observe most Type Ia SNRs with $d_\mathrm{SNR}\le\ 5.0\ \mathrm{kpc}$ and CC SNRs with $d_\mathrm{SNR}\le\ 10.0\ \mathrm{kpc}$ and younger than 5000 yr, provided that they have a particle acceleration efficiency similar to Tycho and RX~J1713. These results are encouraging in that the number of the SNRs whose VHE emission will be detected at a distance $d_\mathrm{SNR}\le\ 5.0\ \mathrm{kpc}$ will dramatically increase in the CTA era.

\section{Conclusion}\label{Conclusion}
In this study, we model the time evolution of SNRs using a hydrodynamical simulation coupling with efficient particle acceleration based on previous works \citep[e.g.,][]{2004APh....21...45B,2010APh....33..307C,2010MNRAS.407.1773C,LEN2012}, and investigate how their broadband non-thermal SEDs evolve in various kinds of ambient environments. We prepare three models for the ambient medium, including a uniform ISM-like case for Type Ia SNRs, a power-law CSM from a steady isotropic stellar wind for CC SNRs, and a case with a pre-SN enhanced mass-loss from a massive star that creates a dense confined CSM shell surrounding the ejecta.

In the Ia models with a uniform ISM, while the $\pi^0$-decay flux increases with time, IC flux does not vary much with its spectral peak shifting to lower energy as the SNR ages. In the CC models with a simple power-law CSM, while $\pi^0$-decay flux decreases with time, the IC contribution increases with time on the contrary. We found that the key aspects that dictate these evolutionary trends are the density distribution of the interaction targets for each emission component, and the rate of energy loss of the electrons due to synchrotron radiation.
In our models, since the interaction target is the ambient gas for $\pi^0$-decay and the uniform CMB radiation field for IC, the spatial distribution of the ambient gas density is a key to understand the evolution of the $\gamma$-ray spectrum, including a possible transition between a leptonic and a hadronic origin at a certain evolutionary stage. Moreover, the accelerated electrons lose their energy via synchrotron radiation due to a highly amplified magnetic field in the uniform ISM cases. 
Our results are consistent with the previously proposed picture that the ISM/CSM gas density decide the dominant component of $\gamma$-ray emission from a SNR \citep[e.g.,][]{2012ApJ...761..133Y}. In addition, we propose that not only the number density of the ambient environment but also the distribution of magnetic field is also important in understanding the time-evolution of VHE emission. In the case of an enhanced mass loss from a massive star progenitor, the production of secondary particles are found to be very efficient in the dense confined CSM shell and contribute importantly to the overall SED. For example, they can dominate the synchrotron radiation after the SNR breaks out from the shell into a tenuous wind. 

A comparison between our models and observations show a broad agreement. A dramatic enlargement of the sample size of $\gamma$-ray emitting SNRs is anticipated in the CTA era to further constrain the parameter space in our systematic survey of SNR broadband models.  CTA will have a sufficient sensitivity to detect VHE emission from most Ia and CC SNRs in various environments with a distance within $\sim 5.0$ kpc. Future observations by CTA will reveal the detailed morphological and spectral properties of $\gamma$-ray emissions from SNRs and make important progress on our understanding of particle acceleration mechanism at astrophysical collisionless shocks. 

We note that the current study has only examined several simple models for the ambient environment, which in reality can be much more complicated such as the presence of a cavity, dense shells, clumpy winds and MCs, etc. Our code is designed to be modular which makes it easy for us to expand into a broader parameter space, including more complicated models for the environment. In future work, we will also explore other important physics such as the acceleration of heavier ions, thermal X-ray line emission, radiative shocks and so on.     

\acknowledgments 
The authors acknowledge important discussions with K. Maeda concerning this work. SHL acknowledges support from the Kyoto University Foundation.

\bibliographystyle{aa} 
\bibliography{reference}

\begin{deluxetable*}{lllllll}
\centering
\tablecolumns{6}
\tablewidth{17cm}
\tablecaption{SED references} 
\tablehead{
SNR & radio & X-ray & GeV $\gamma$-ray & TeV $\gamma$-ray 
}
\startdata
G34.7$-$0.4 & - & - & \cite{2013Sci...339..807A} & - 
\\
&&& \cite{2011ApJ...742L..30G} & 
\\
\hline
G111.7$-$2.1 & \cite{1967AZh....44..984A} &\cite{2009PASJ...61.1217M} & \cite{2010ApJ...710L..92A} & \cite{2010ApJ...714..163A} 
\\
&& \cite{2016ApJ...825..102W} && \cite{2015BLPI...42..169S}
\\
&&&  & \cite{2017MNRAS.472.2956A}
\\
\hline
G120.1$+$1.4 & \cite{2006AA...457.1081K}& \cite{2014ApJ...797L...6T} & \cite{2012ApJ...744L...2G}& \cite{2011ApJ...730L..20A} 
\\
&&&\cite{2017ApJ...836...23A} & \cite{2017ApJ...836...23A}
\\
\hline
G189.1$+$3.0 & && \cite{2010ApJ...710L.151T}& \cite{2007ApJ...664L..87A}
\\
&&&\cite{2013Sci...339..807A}&\cite{2009ApJ...698L.133A}
\\
\hline
G260.4$-$3.4 & \cite{2012ApJ...759...89H} & & \cite{2017ApJ...843...90X}&\cite{2015AA...575A..81H}
\\
&\cite{2016AA...586A.134P}
\\
\hline
G266.2$-$1.2 & \cite{2000AA...364..732D} && \cite{2011ApJ...740L..51T} & \cite{2018AA...612A...7H}
\\
\hline
G327.6$+$14.6 & \cite{2001ApJ...558..739A}& \cite{2008PASJ...60S.153B} & \cite{2017ApJ...851..100C}& \cite{2010AA...516A..62A}
\\
\hline
G347.3$-$0.5 & \cite{2004ApJ...602..271L} & \cite{2008ApJ...685..988T}& \cite{2011ApJ...734...28A}& \cite{2007AA...464..235A}
\\
&\cite{2009AA...505..157A}&&&\cite{2011AA...531C...1A}
\\
&&&&\cite{2018AA...612A...6H}
\enddata
\tablecomments{The SED data references in each wavelength in Fig.~\ref{graph:SED}, Fig.~\ref{test:tycho}, and Fig.~\ref{test:RXJ1713}.}
\label{table:data}
\end{deluxetable*}

\begin{deluxetable*}{l|cccccccc}
\centering
\tablecolumns{9}
\tablewidth{15cm}
\tablecaption{Model parameter} 
\tablehead{
Model & $M_\mathrm{ej}$ & $n_\mathrm{ISM}$ & $\dot{M}$ & $V_\mathrm{w}$ & $\dot{M_2}$ & $V_\mathrm{w,2}$ & $\chi_\mathrm{inj}$\\
  & [$M_\odot$] & [cm$^{-3}$] & [$M_\odot$ yr$^{-1}$] & [km s$^{-1}$] & [$M_\odot$ yr$^{-1}$] & [km s$^{-1}$] &  
}
\startdata
A0\tablenotemark{a} & 1.4 & 0.3 & - & - & - & - & 3.6 & \\
A1 & 1.4 & 0.01 & - & - & - & - & 3.6 & \\
A2 & 1.4 & 0.1 & - & - & - & - & 3.6 & \\
A3 & 1.4 & 1.0 & - & - & - & - & 3.6 & \\
\hline
B0\tablenotemark{b} & 3.0 & - & 7.5$\times$10$^{-6}$ & 20 & - & - & 3.75 & \\
B1 & 3.0 & - & 1.0$\times$10$^{-6}$ & 20 & - & - & 3.75 & \\
B2 & 3.0 & - & 1.0$\times$10$^{-5}$ & 20 & - & - & 3.75 & \\
B3 & 3.0 & - & 1.0$\times$10$^{-4}$ & 20 & - & - & 3.75 & \\
B4 & 10.0 & - & 1.0$\times$10$^{-6}$ & 20 & - & - & 3.75 & \\
B5 & 10.0 & - & 1.0$\times$10$^{-5}$ & 20 & - & - & 3.75 & \\
B6 & 10.0 & - & 1.0$\times$10$^{-4}$ & 20 & - & - & 3.75 & \\
\hline
C\tablenotemark{c} & 10.0 & - & 5.0$\times$10$^{-6}$ & 15 & 0.01 & 1000 & 3.75
\enddata
\tablenotetext{a}{All model A use an exponential profile for the ejecta,  $E_\mathrm{SN}$ = 10$^{51}$ erg, $T_0=10^4\ \mathrm{K}$, $B_0=4.0\ \mu\mathrm{G}$, and $d_\mathrm{SNR}$ = 3.2 kpc.}
\tablenotetext{b}{All model B use a power-law profile for the ejecta with $n_\mathrm{pl}$ = 7, $E_\mathrm{SN}$ = 10$^{51}$ erg,  $T_0=10^4\ \mathrm{K}$, $\sigma_\mathrm{w}=0.004$, and $d_\mathrm{SNR}$ = 1.0 kpc.}
\tablenotetext{c}{This model uses a power-law profile for the ejecta with $n_\mathrm{pl}$ = 7, $E_\mathrm{SN}$ = 10$^{51}$ erg, $T_0=10^4\ \mathrm{K}$, $n_\mathrm{pl,2}$ = 1.5, and $d_\mathrm{SNR}$ = 1.0 kpc.}
\label{table:param}
\end{deluxetable*}

\begin{turnpage}
\begin{deluxetable*}{llccccccccc}
\centering
\tablecolumns{11}
\tablewidth{25cm}
\tablecaption{Observation data} 
\tablehead{
SNR & common name & type\tablenotemark{a} & age & distance & radius & velocity & F$_\mathrm{1GHz}$ & F$_\mathrm{1-100GeV}$\tablenotemark{b} & F$_\mathrm{1-10TeV}$\tablenotemark{c} & Refs.\\
& &  & [yr] & [kpc] & [deg] & [$''$/yr] & [Jy] & [10$^{-9}$ cm$^{-2}$ s$^{-1}$] &  [10$^{-13}$ cm$^{-2}$ s$^{-1}$] &  
}
\startdata
G1.9$+$0.3 & & Ia & 150 - 220 & 8.5 & 0.01 & 0.32$\pm$0.02 & 0.6 & 0.27 & 0.72 & 
[1]\\
G4.5$+$6.8 & Kepler & Ia & 414 & 2.9 - 4.9 & 0.03 & 0.23$\pm$0.01 & 19 & 0.65 & 0.12 & [2]
\\
G15.9$+$0.2 & & CC & 1000 - 3000 & 8.5 & 0.05 & 0.021$\pm$0.001 & 5.0 & 1.6 & 2.6 & [3]
\\
G34.7$-$0.4 & W44 & CC & 7900 - 8900 & 2.7 - 3.3 & 0.31 $\pm$ 0.02 & 0.047 & 240 & 54.95 $\pm$ 2.68 & 11.2 & [4]
\\
G43.3$-$0.2 & W49B & CC & 2900 - 6000 & 10.9 - 11.7 & 0.07 & 0.022 & 38 & 19.24 $\pm$ 1.01 & - & [5]
\\
G67.7$+$1.8 & & CC & 5000 - 13000 & 7.0 - 17.0 & 0.11 & - & 1.0 & 0.43 & 15.3 & [6]
\\
G111.7$-$2.1 & Cas A & CC & 316 - 352 & 3.3 - 3.7 & 0.041 $\pm$ 0.001 & 0.31 $\pm$ 0.02 & 2400 & 6.25 $\pm$ 0.42 & 5.8 $\pm$ 1.2 & [7]
\\
G120.1$+$1.4 & Tycho, SN1572 & Ia & 446 & 2.4 - 5.0 & 0.07 & 0.30 $\pm$ 0.10 & 56 & 1.06 $\pm$ 0.33 & 1.1 $\pm$ 0.4 & [8]
\\
G189.1$+$3.0 & IC443, Jellyfish Nebula & CC & 3000 - 30000 & 0.7 - 2.0 & 0.33 $\pm$ 0.01 & 0.012 & 165 & 57.27 $\pm$ 1.15 & -  & [9]
\\
G260.4$-$3.4 & Puppis A & CC & 3700 - 4500 & 1.3 - 2.2 & 0.33 $\pm$ 0.02 & 0.13 & 130 & 8.04 $\pm$ 0.56 & - & [10]
\\
G266.2$-$1.2 & Vela Jr, RX J0852.0-4622 & CC & 2400 - 5100 & 0.5 - 1.0 & 1.19 $\pm$ 0.04 & 0.35 $\pm$ 0.13 & 50 & 12.11 $\pm$ 0.89 & 200 $\pm$ 6.58 & [11] 
\\
G272.2$-$3.2 & & Ia & 3600 - 11000 & 2.0 - 10.0 & 0.12 & 0.036 & 0.4 & 1.2 & 5.6 & [12]
\\
G291.0$-$0.1 & & CC & 1300 - 10000 & 3.5 - 6.0 & 0.23 & - & 16 & 13.04 $\pm$ 0.78 & 4.9 & [13]
\\
G292.0$+$1.8 & & CC & 2930 - 3050 & 6.0 & 0.2 & 0.043 & 15 & 6.17 $\pm$ 0.43 & 3.2 & [14]
\\
G296.1$-$0.5 & & CC & 2800 - 28000 & 2.0 - 4.0 & 0.25 & 0.040$\pm$0.003 & 8 & 2.0 & 3.6 & [15]
\\
G306.3$-$0.9 & & Ia & 1300 - 4600 & 8.0 & 0.02 & 0.018$\pm$0.001 & 0.16 & 1.1 & 1.1 & [16]
\\
G308.4$-$1.4 & & CC & 5000 - 7500 & 9.1 - 10.7 & 0.07 & 0.016$\pm0.004$ & 0.4 & 0.58 & 3.5 & [17]
\\
G309.2$-$0.6 & & CC & 700 - 4000 & 2.0 - 6.0 & 0.11 & - & 7 & 0.96 & 3.8 & [18]
\\
G315.4$-$2.3 & RCW86 & Ia & 2000 - 10000 & 2.3 - 3.2 & 0.35 & 0.13$\pm$0.06 & 49 & 34 & 18.2 $\pm$ 9.4& [19]
\\
G327.6$+$14.6 & SN1006 & Ia & 1012 & 1.6 - 2.2 & 0.25 & 0.29 & 19 & 0.72 & 3.7 $\pm$ 0.8 & [20]
\\
G330.2$+$1.0 & & CC & 1000 - 3000 & $\ge$5.0 & 0.08 & 0.20 & 5 & 0.64 & 7.3 & [21]
\\
G347.3$-$0.5 & RX J1713.7-3946 & CC & 1625 & 1.0 & 0.53 $\pm$ 0.03 & 0.82 $\pm$ 0.06 & 30 & 4.94 $\pm$ 0.81 & 146 $\pm$ 6 & [22]
\\
G350.1$-$0.3 & & CC & 600 - 1200 & 4.5 - 9.0 & 0.03 & 0.11$\pm$0.03 & 6 & 3.3 & 1.6 & [23]
\enddata
\tablecomments{The SNR observation data, radio flux from \cite{2017yCat.7278....0G}, GeV flux from \cite{2016ApJS..224....8A}, TeV flux from \cite{2018AA...612A...6H}, and the other data from {\it SNRcat} (available at "http://www.physics.umanitoba.ca/snr/SNRcat/") and each references; [1] \cite{2011ApJ...737L..22C}, \cite{2014MNRAS.441..790H}, [2] \cite{2008AA...488..219A}, \cite{2008ApJ...689..231V}, [3] \cite{2018MNRAS.479.3033S}, [4] \cite{2012PASJ...64..141U}, \cite{2013Sci...339..807A}, [5] \cite{2007ApJ...654..938K}, \cite{2014ApJ...793...95Z}, [6] \cite{2009AA...494.1005H}, [7] \cite{2003ApJ...589..818D}, [8] \cite{2010ApJ...709.1387K}, \cite{2012ApJ...744L...2G}, [9] \cite{2013Sci...339..807A}, \cite{2017MNRAS.472...51A}, [10] \cite{2017MNRAS.464.3029R}, [11] \cite{2015ApJ...798...82A}, \cite{2018AA...612A...7H}, [12] \cite{2013AA...552A..52S}, [13] \cite{1986MNRAS.219..815R}, [14] \cite{2003ApJ...594..326G}, \cite{2003ApJ...583L..91G}, [15] \cite{2012MNRAS.419.1603G}, [16] \cite{2013ApJ...766..112R}, \cite{2017MNRAS.466.3434S}, [17] \cite{2012AA...544A...7P}, [18] \cite{2001ApJ...548..258R}, [19] \cite{2013MNRAS.435..910H}, \cite{2018AA...612A...4H}, [20] \cite{2013Sci...340...45N}, \cite{2017ApJ...851..100C}, [21] \cite{2014MNRAS.441..790H}, \cite{2018ApJ...855..118W}, [22] \cite{2016PASJ...68..108T}, \cite{2018AA...612A...6H}, [23] \cite{2011ApJ...731...70L}.}
\tablenotetext{a}{type of SN explosion, `Ia' implies thermonuclear explosion and `CC' implies core-collapse SN. About the SNRs whose explosion type is not known much, it is assumed to be `CC' in this table.}
\tablenotetext{b}{the integrated flux by {\it Fermi}-LAT in 1-100 GeV range, but those without error are the 99 $\%$ confidence upper limit with photon index $\Gamma = 2.5$ because the SNR is not observed by {\it Fermi}-LAT, see \cite{2016ApJS..224....8A} for details.}
\tablenotetext{c}{the integral flux by H.E.S.S. in 1-10 TeV range, that with error is the detected value and that without error is the upper limit of 99 $\%$ confidence level with $\Gamma = 2.3$.}
\label{table:data}
\end{deluxetable*}
\end{turnpage}

\end{document}